\documentclass[a4paper,11pt]{article}
\pdfoutput=1

\usepackage{jcappub}
\usepackage[T1]{fontenc} 
\usepackage[utf8]{inputenc}
\usepackage[T1]{fontenc} 
\usepackage{natbib, hyperref}
\usepackage[varg]{txfonts}
\usepackage{caption}
\usepackage{comment}
\usepackage{blindtext}
\usepackage{autobreak}
\usepackage{tikz}
\usetikzlibrary{matrix}
\usepackage{bm}
\usepackage{amsmath}
\usepackage{float}
\usepackage{color}
\usepackage{xcolor}
\usepackage{appendix}
\usepackage{hyperref}
\usepackage{comment}
\usepackage{multirow}  
\usepackage[varg]{txfonts}
\usepackage{graphicx}

\newcommand{\Hbb}{\mathbb{H}}
\newcommand{\Qbb}{\mathbb{Q}}
\newcommand{\Rbb}{\mathbb{R}}
\newcommand{\Jbb}{\mathbb{J}}
\newcommand{\Sbb}{\mathbb{S}}

\newcommand{\Xbb}{\mathbb{X}}
\newcommand{\Xbba}{\mathbb{X}^{(1)}}
\newcommand{\Xbbb}{\mathbb{X}^{(2)}}
\newcommand{\Xbbc}{\mathbb{X}^{(3)}}
\newcommand{\Xbbd}{\mathbb{X}^{(4)}}

\newcommand{\peq}{\,\stackrel{\text{pstf}}{=}\,}

\newcommand{\dd}{\text{d}}

\title{Expanding covariant cosmography of the local Universe: incorporating the snap and axial symmetry}

\author[]{Basheer Kalbouneh$^{1}$, Jessica Santiago$^{2}$,  Christian Marinoni$^{1}$,\\  Roy Maartens$^{3,4,5}$, Chris Clarkson$^{6,3}$, Maharshi Sarma$^{1}$}

\affiliation[1]{\small{Aix Marseille Univ, Universit\'e de Toulon, CNRS, CPT, Marseille, France}}
\affiliation[2]{\small{Leung Center for Cosmology and Particle Astrophysics,
National Taiwan University, Taiwan}}
\affiliation[3]{\small{Department of Physics \& Astronomy, University of the Western Cape, Cape Town 7535, South Africa}}
\affiliation[4]{\small{Institute of Cosmology \& Gravitation, University of Portsmouth, Portsmouth PO1 3FX, United Kingdom}}
\affiliation[5]{\small{National Institute for Theoretical \& Computational Sciences, Cape Town 7535, South Africa}}
\affiliation[6]{\small{Department of Physics \& Astronomy, Queen Mary University of London, London E1 4NS, United Kingdom}}
\abstract{
Studies show that the model-independent, fully non-perturbative covariant cosmographic approach is suitable for analyzing the local Universe $(z\lesssim 0.1)$. However, accurately characterizing large and inhomogeneous mass distributions requires the fourth-order term in the redshift expansion of the covariant luminosity distance $d_L(z,\boldsymbol{n})$.
We calculate the covariant snap parameter $\Sbb$ and its spherical harmonic multipole moments using the matter expansion tensor and the evolution equations for lightray bundles.
The fourth-order term adds 36 degrees of freedom, since the highest independent multipole of the snap is the 32-pole (dotriacontapole) ($\ell=5$).
Including this term helps to de-bias estimations of the covariant deceleration parameter.  Given that observations suggest axially symmetric anisotropies in the Hubble diagram for $z \lesssim 0.1$ and theory shows that only a subset of multipoles contributes to the signal, we demonstrate that only 12 degrees of freedom are needed for a model-independent description of the local universe. We use an analytical axisymmetric model of the local Universe,  with data that matches the Zwicky Transient Facility survey, in order to provide a numerical example of the amplitude of the snap multipoles and to forecast precision.}

\makeatletter
\gdef\@fpheader{{Published at JCAP}\\{DOI: 10.1088/1475-7516/2025/02/076}}
\makeatother

\begin{document}
\maketitle
\flushbottom

\section{Introduction}

Recent challenges to the cosmological principle (CP), including tensions in the estimation of the Hubble constant $H_0$ 
\cite{Schwarz:2007wf, Kashlinsky:2008ut, Antoniou:2010gw, Cai:2011xs, Kalus:2012zu, Wang:2014vqa, Yoon:2014daa, Tiwari:2015tba, Javanmardi2015, Bengaly:2015nwa, Colin:2017juj, Rameez:2017euv, Migkas:2020fza, Migkas:2021zdo, Secrest:2020has, Siewert:2020krp, Luongo:2021nqh, Krishnan:2021jmh, Sorrenti:2022zat, Aluri:2022hzs, Cowell_Dhawan_Macpherson_2023, Hu:2023eyf, Dainotti:2021pqg,Peebles:2022akh,Sorrenti:2024ztg}, 
highlight the need for a fully covariant, model-independent and non-perturbative investigation of the geometry of the Universe.

In a series of papers, \cite{Maartens:2023tib} (paper I) and \cite{Kalbouneh:2024}  (paper II), we explored a more comprehensive way to characterize the anisotropic cosmic expansion rate in the local Universe, where the CP  does not apply.
In this covariant cosmographic framework, redshift and distance measures are directly related to the kinematical properties of the matter flow, bypassing the need for cosmological perturbation theory. This approach directly probes the spacetime geometry without assuming a smooth background or relying on the model-dependent concept of peculiar velocities to describe deviations from such a hypothetical background.

The structural information about the cosmic metric is encapsulated in a set of {\it covariant cosmographic} parameters, particularly within the finite set of its multipoles. 
Their  estimation using observational data began shortly after their initial introduction by Kristian and Sachs \cite{Kristian:1965sz}. 
In 1967, Trendowski \cite{Trendowski1967}, by analyzing the  galaxy catalog of Humason et al. \cite{1956AJ.....61...97H} found that the ratio of the quadrupole of the covariant Hubble parameter $\mathbb{H}$  to its monopole cannot exceed $\sim10\%$. However, early studies were hindered by limited data and did not progress beyond constraining the lowest-order term in the redshift expansion of the distance function.

Recently, several studies have reported preliminary results on an expanded set of cosmographic parameters, leveraging new and richer datasets.
For example, \cite{Colin:2019opb} used the JLA dataset \cite{SDSS:2014iwm}, while \cite{Dhawan} utilized both JLA and the Pantheon \cite{Pan-STARRS1:2017jku} samples. In \cite{Cowell:2022ehf}, the authors used the Pantheon+ dataset \cite{Scolnic:2021amr}. Apart from the monopoles, these studies focus on the quadrupole of $\Hbb$ and the dipole of $\Qbb$,  modifying both by multiplying them by a factor that decays with redshift. They found that the quadrupole of $\Hbb$ is consistent with zero for both the JLA and Pantheon datasets. However, in Pantheon+, a deviation of about $3\%$ from zero was observed, reaching a significance level of $2\sigma$. For the dipole of $\Qbb$, the signal is either consistent with or significantly higher than zero and can be found either in the direction of the CMB dipole or in the opposite direction, depending on the frame used for redshift and distance measurements (heliocentric or CMB).

Regarding disagreements in the literature concerning the dipole estimations, in paper I and paper II, we have uncovered the effects of the local boost  of the observer  relative  to the matter rest frame  on the measurement of cosmographic parameters, provided a covariant method to disentangle the kinematic effects due to the observer's motion  from intrinsic anisotropies and shown the imprint of the frame choice adopted in the results obtained and the importance of disentangling  these effects. 
This is due to the fact that, although the general expansion of the covariant luminosity distance is valid only for an observer comoving with the cosmological matter frame (which we are not), one can also consider the frame of an observer boosted with respect to the matter frame -- as long as the observer's boost velocity is properly taken into account as a new degree of freedom to be determined. This consideration is essential for estimating the covariant cosmographic parameters in an unbiased manner.  

Here, in paper III, we constrain the  kinematics of matter around the observer using the covariant cosmographic analysis of the  luminosity distance \cite{Maartens:2023tib,Kalbouneh:2024}  (see also
\cite{Kristian:1965sz, MacCallum_Ellis_1970,ellis_2009,ellis_1983,ellis85, Hasse:1999,Clarkson_theses_2000,clarkson_maartens_2010,Umeh:2013,heinesen_2021}). 
Discarding the isotropy assumption, the covariant cosmographic parameters $\mathbb{H}, \mathbb{Q}, \mathbb{J}$ and $\mathbb{S}$  are  {line-of-sight dependent functions}, which describe the local structure of the cosmic spacetime without assuming a specific background metric and without the need for peculiar velocity corrections.

\begin{figure*}
  \centering
  \includegraphics[scale=0.4]{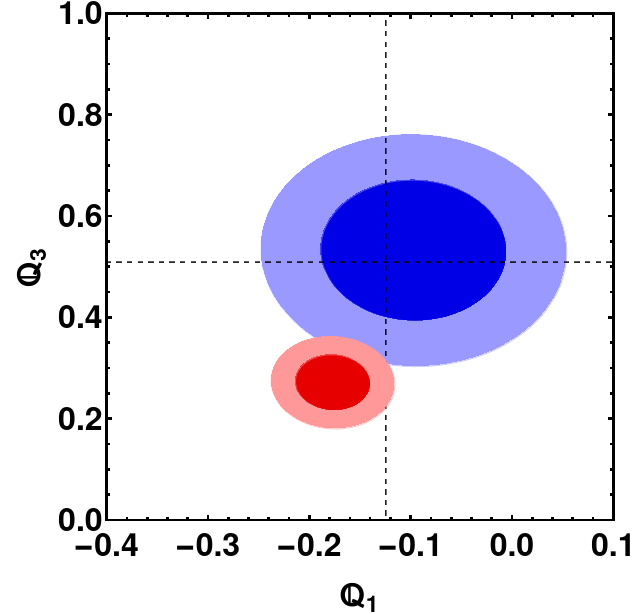}
  \caption{Likelihood of the dipole $\mathbb{Q}_1$ and octupole $\mathbb{Q}_3$  of the deceleration parameter for model $M2$ (from \cite{Kalbouneh:2024} and \S\ref{sec_num} here). Red indicates the likelihoods of $\mathbb{Q}_1$  and $\mathbb{Q}_3$ without including the snap $\mathbb{S}$ in the fit;  blue shows the likelihoods with the snap included. Dashed lines intersect at the true input values.}
  \label{HABC_1}
\end{figure*}

In paper II \cite{Kalbouneh:2024},  using a perturbative model motivated by observational evidence, we forecasted that future local Universe data, like that from the Zwicky Transient Facility (ZTF) survey, could determine the amplitude of the lower multipoles (up to $\ell=3$) of the covariant deceleration parameter $\mathbb{Q}_o$ at the observer's position. However, the estimation of the dipole and octupole are biased if the luminosity distance expansion is truncated at third order in redshift. A major result was that including the often-overlooked local motion of the observer relative to matter as a free parameter helps to debias the dipole estimate, but does not correct the systematic shift of the octupole. Indeed, in this expansion approximation,  the octupoles of the third-order covariant cosmographic parameters, the jerk $\Jbb_o$ and the curvature $\Rbb_o$, which are systematically multiplied by $z^3$, provide small contributions and do not improve the likelihood analysis. In this paper III, we tackle this issue by calculating the fourth-order covariant cosmographic parameter, the snap $\mathbb{S}$. Our key result is that, as expected,  including $\mathbb{S}$ effectively debiases the estimated amplitude of the dipole and octupole of the deceleration parameter $\mathbb{Q}_o$ -- as shown in Figure \ref{HABC_1}.

The paper is structured as follows. In Section  \ref{S:FLRW} we briefly review the cosmographic approach in FLRW spacetimes.  Section \S\ref{S: obtaining da(z,n)}  introduces the foundations of the covariant cosmographic approach and presents for the first time the explicit calculation of the covariant snap function in a general spacetime. In Section~\ref{sec_dof}, we present the degrees of freedom of the distance-redshift relation at different orders in redshift. In Section~\ref{sec_num}, we use a simple yet realistic analytical axisymmetric model of mass inhomogeneities in the local Universe, motivated by observational evidence, to forecast the amplitude of the snap multipoles. We generate local Universe data ($z<0.1$), based on this model, using the redshift distribution and the uncertainties in measured distances as given by the ZTF survey, in order to forecast the precision of future measurements. The multipoles of the covariant jerk $\mathbb{J}$ and snap $\mathbb{S}$ parameters, together with the associated curvature terms, are presented in Section~\ref{S: obtaining da(z,n)} in terms of the covariant functions $\mathbb{X}^{(3,4)}$ in \eqref{X3}, \eqref{X4} and $\mathbb{Y}^{(1,2)}$ in \eqref{y12}. Their multipoles, written in terms of the kinematic properties of the matter, are presented in Appendix \ref{S:Appendix}.

\enlargethispage{0.5cm}

Hereafter we adopt the Einstein summation convention for repeated indices. Latin indices $a,b,\cdots$ indicate arbitrary coordinates or tetrads in a generic spacetime; Greek indices $\mu,\nu,\cdots$ refer to a choice of coordinates in a specific spacetime. We use natural units ($c=1$) and the metric signature is $(-+++)$.

\section{Expansion of the angular diameter distance in FLRW}
\label{S:FLRW}
The CP leads to the Friedmann–Lemaître–Robertson–Walker metric (FLRW) which, expressed in comoving (spherical) coordinates, is given by:
\begin{equation}
\dd s^{2}= g_{\mu\nu}\;\dd x^\mu \dd x^\nu =
-\dd t^{2}+a^{2}(t)\left[\frac{\dd r^{2}}{1-K r^2} + r^{2}\Big(\dd\theta^{2} + \sin^{2}\theta\, \dd\phi^{2}\Big) \right],
\label{FLRWmetric1}
\end{equation}
where $t$ is the coordinate time (also called cosmic time),   $a(t)$ is the scale factor of the universe, and $K$ describes the only three possible uniform spatial sections that are compatible with the cosmological principle. 
By rescaling $K, \; r$ and $a(t)$, one can choose $K$ to take one of the values $(-1,0,1)$, which are related to open (negatively curved), flat, and closed (positively curved) spaces respectively. In this case, the scale factor will have the dimension of the distance.

The angular diameter distance $d_A$ is defined
for a geodesic bundle converging at the point of observation, as the square root of the ratio between the transverse physical area of an object and the observed solid angle \cite{Maartens:2023tib}
\begin{equation}
    d_A\equiv\sqrt{\frac{\dd A}{\dd\Omega}}.
\label{dAdef1}
\end{equation}
In an FLRW model, the angular diameter distance measured by an observer today (at $t=t_0$) for an emitter at $t=t_e$, is
\begin{equation}
d_{A}=a(t_e)\; S_{K}\left(\int^{t_{0}}_{t_{e}}\frac{dt'}{a(t')}\right),
\end{equation}
where $S_{K}(\chi)$ is the coordinate distance travelled by the photon, and it is defined for different geometries as:
\begin{align}
S_{K}(\chi)=\begin{cases} 
     \, \sin(\chi) &,\;\; K=+1  \\
    \, \chi &,\;\; K=0 \\
     \, \sinh(\chi) &,\;\; K=-1
   \end{cases} \quad.
\end{align}

The redshift  $z$ is related to the ratio between the energy of the photon at emission $E_e$ and its observed energy $E_0$ by 
\begin{equation}
1+z\equiv \frac{E_e}{E_0}=\frac{a(t_0)}{a(t_e)}\;,
\end{equation}
which allows us to rewrite the angular diameter distance in terms of the redshift as:
\begin{equation}
d_{A}=\frac{a_{0}}{1+z}S_{K}\left(\frac{1}{a_{0}}\int^{z}_{0}\frac{dz'}{H(z')}\right)\;.
\label{dAeqn1}
\end{equation}
Here, $H\equiv \dot a/a$ is the standard Hubble parameter, with the overdot denoting a derivative with respect to  cosmic time. By expanding relation (\ref{dAeqn1}) around $z=0$, one can find \cite{Visser:2003vq}
\begin{align}
    d_A(z)&=\frac{z}{H_0}\bigg[1-\frac{1}{2}\left(3+q_0\right)z+\frac{1}{6}\left(11+7q_0+3q_0^2+\Omega_{K0}-j_0\right)z^2
     \label{daz2}\\ \notag
    &\;\;\;\;+\frac{1}{24}\left(-50-10\Omega_{K0}-q_0(46+6\Omega_{K0}+39 q_0+15 q_0^2)+j_0(13+10q_0)+s_0\right)z^3\bigg]+\mathcal{O}(z^5),
\end{align}
where
\begin{subequations}
\begin{align}
    q_0&\equiv-\frac{\ddot{a}_0}{a_0 H_{0}^2}=\frac{1}{2}\Omega_{m0}-\Omega_{\Lambda0} &&\text{deceleration}, \\ 
    j_0&\equiv\frac{\dddot{a}_0}{a_0 H_{0}^3}=\Omega_{m0}+\Omega_{\Lambda 0}                    &&\text{jerk},
\\
\Omega_{K0}&\equiv\frac{-K}{a_0^2 H_{0}^2}=1-\Omega_{m0}-\Omega_{\Lambda 0}                    &&\text{curvature},
\\
s_0&\equiv\frac{\ddddot{a}_0}{a_0 H_{0}^4}=\frac{1}{2}\Omega_{m0}\big(\Omega_{\Lambda0}-\Omega_{m0}-6\big)+\Omega_{\Lambda0}^2                     &&\text{snap}.
\end{align}
\end{subequations}
In order to obtain the second equalities above, one needs to make use of the Einstein field equations, which reduce to the Friedmann equations in FLRW.
Note also that $\Omega_{m}={8\pi G \rho_m}/{(3H^2)}$ and $\Omega_{\Lambda}={\Lambda}/{(3H^2)}$ are the dimensionless density parameters, $\rho_m$ is the matter density as observed by comoving observers, and $\Lambda$ is the cosmological constant.

The expansion of the luminosity distance can be obtained by using Etherington's reciprocity theorem $d_L=(1+z)^2 d_A$, which gives us: 
\begin{align}
    d_L(z)&=\frac{z}{H_0}\bigg[1+\frac{1}{2}\left(1-q_0\right)z +\frac{1}{6}\left(3q_0^2+q_0-1+\Omega_{K0}-j_0\right)z^2
    \nonumber
    \\
    &\;\;\;\;+\frac{1}{24}\left(2-q_0(2+15q_0+15q_0^2)+5j_0(1+2q_0) -2\Omega_{K0}(1+3q_0)+s_0\right)z^3\bigg]+\mathcal{O}(z^5) \;.
\end{align}

\section{Expansion of the angular diameter distance in a general spacetime}
\label{S: obtaining da(z,n)}

\subsection{Null geodesic congruence}

Consider the time-like congruence defined by the 4-velocity field $u^a$  comoving with the matter distribution, i.e. the matter frame.
Following \cite{Maartens:2023tib,Kalbouneh:2024}, we assume that the matter distribution is  described by a pressureless `dust' fluid, so that $u^b\nabla_b u^a =0$. We further assume that it is irrotational, i.e.  $\omega_{ab} = h_{c [a} h_{b] d} \nabla^c u^d =0$, where $h_{ab}=g_{ab}+u_au_b$ projects into the rest-space of $u^a$.
A matter observer $O$, comoving with $u^a$, receives information via light rays, which are described by a null geodesic congruence,  defined by the affinely parametrized tangent vector $p^a$, which is irrotational since it generates the past lightcones \cite{Ellis:2012}. 
We can decompose it as
\begin{equation}
    p^a=E\big(u^a-n^a\big) \quad\mbox{where} \quad E= -p^a u_a\,,~~ n^a n_a=1\,,~~ n^a u_a=0 \;.
\end{equation}
Here  $E$ is the photon energy, as measured by $O$, and $n^a$ is the unit  direction of the incoming light ray. The screen-space projection tensor \cite{Sachs:1961zz,Ellis:2012},
\begin{equation}
    S^{ab}=h^{ab}-n^a n^b \quad\mbox{where} \quad S^{ab}u_a=0=S^{ab}n_a\,,
    \label{E: screenproj}
\end{equation}
projects into the observer's  2-plane (screen) that is orthogonal to the lightray direction. 
Using \eqref{E: screenproj}, the expansion tensor for the null congruence can be decomposed as 
\begin{equation}
    \hat\Theta_{ab}\equiv S^c_b S^d_a \;\nabla_c p_d= 
    \frac{1}{2}
    \hat\Theta \, S_{ab}+\hat{\sigma}_{ab}\,,
\end{equation}
where 
\begin{equation}
    \hat\Theta\equiv 
    S^{ab}\;\nabla_a p_b=
    \nabla_a p^a \qquad\text{and}\qquad
    \hat\sigma_{ab}\equiv \hat\Theta_{ab}-\frac{1}{2}\hat\Theta \,S_{ab}\,,
\end{equation}
are the area expansion rate (the trace) and the 
shear rate (symmetric traceless part)  of $\hat\Theta_{ab}$, respectively. 
For our purpose, we only need the evolution of $\hat\Theta$ 
along the lightrays, which can be derived using the trace of the geodesic  deviation equation \cite{Sachs:1961zz,Kristian:1965sz} (see also \cite{Wald:1984rg,Carroll:2004st,Poisson:2009pwt,Ellis:2012,Clarkson_ellis_roy_2012,Fleury:2015hgz}):
\begin{align}
    p^a\nabla_a \hat\Theta+\frac{1}{2}\hat\Theta^2+
    \hat\sigma_{ab}\hat\sigma^{ab}
    =-R_{ab}p^ap^b,
   \label{eqn_scalars1}
\end{align}
where  $R_{ab}$ is the Ricci tensor.

\subsection{Angular diameter distance expansion} \label{S: distance expansion}

We start with the angular diameter distance $d_A$. Assuming that it is differentiable and not multi-valued, we can expand it in terms of the affine parameter $\lambda$, where $\dd /\dd\lambda \equiv p^a\nabla_a$, around the event of observation $o$:
\begin{equation}
    d_A=d_A \big|_o+\frac{\dd d_A}{\dd\lambda}\bigg|_o\lambda+ \frac{1}{2}\frac{\dd^2 d_A}{\dd\lambda^2}\bigg|_o \lambda^2+ \frac{1}{6}\frac{\dd^3 d_A}{\dd\lambda^3}\bigg|_o \lambda^3+ \frac{1}{24}\frac{\dd^4 d_A}{\dd\lambda^4}\bigg|_o \lambda^4+\mathcal{O}(\lambda^5)\;.
\label{dAlambda1}
\end{equation}
Clearly $d_A|_o=0$ and since the spacetime is close to Minkowski near $o$, we have \cite{Maartens:2023tib,heinesen_2021} 
\begin{equation}\label{daic}
    \frac{\dd d_A}{\dd\lambda}\bigg|_o=-E_o \,, \qquad     {\hat{\sigma}_{ab}\big|_o=0}\,.
\end{equation}
Since $\hat\Theta$ is the rate of change of the area of the ray bundle, we have  $\hat\Theta=[\dd (\delta A)/\dd\lambda]/\delta A$. Then $\dd \delta\Omega/\dd \lambda=0$ implies, using  the definition \eqref{dAdef1}, that
\begin{equation}\label{dda}
    \frac{\dd d_A}{\dd\lambda}=\frac{1}{2}d_A \hat\Theta\;.
\end{equation}
Differentiating and using equation (\ref{eqn_scalars1}), we find the  focusing equation 
\begin{equation}\label{ddda}
    \frac{\dd^2 d_A}{\dd\lambda^2}=-\frac{1}{2} d_A\left( 
    \hat\sigma_{ab}\hat\sigma^{ab}
    +R_{ab}p^ap^b\right).
\end{equation}
Therefore $\dd^2 d_A/\dd\lambda^2 \circeq 0$, where $\circeq$ indicates that all the quantities are evaluated at the event of observation $o$. By differentiating \eqref{ddda} and using \eqref{dda}, we find
\begin{align}
    \frac{\dd^3 d_A}{\dd\lambda^3}\circeq 
    \frac{1}{2} E\,R_{ab}p^ap^b 
    \,.
\end{align}
A further differentiation leads to
\begin{equation}
    \frac{\dd^4 d_A}{\dd\lambda^4}\circeq 
   - \frac{\dd d_A}{\dd\lambda}\,p^c\nabla_c \left( 
    \hat\sigma_{ab}\hat\sigma^{ab}
    +R_{ab}p^ap^b\right) \circeq E p^ap^bp^c\nabla_c R_{ab}\,,
\end{equation}
where we used $p^c\nabla_c p^a=0$ and $\hat\sigma^{ab} p^c\nabla_c 
    \hat\sigma_{ab}\circeq 0$ [by \eqref{daic}].
Substituting these results in  (\ref{dAlambda1}), we have
\begin{equation}
    d_A=-E_o\lambda+\frac{1}{12} E_o\left(R_{ab}p^ap^b 
    \right)_o\, \lambda^3+ \frac{1}{24} E_o \left(p^ap^bp^c\nabla_c R_{ab}\right)_o\,\lambda^4+\mathcal{O}(\lambda^5\big).
\label{dAlambda2}
\end{equation}
This is consistent with \cite{Kristian:1965sz} [their eq. (26)].

Following  \cite{Maartens:2023tib,Kalbouneh:2024}, it is useful to use the reverse (past-pointing) and normalised 4-vector $K^a= p^a/(p^b u_b)$, which also removes the $E_o$ factor.
This means that the new affine parameter is $\tilde\lambda=-E_o\lambda$ and $p^a\circeq-E_o K^a$.
Then
\eqref{dAlambda2} becomes
\begin{equation}
    d_A=\tilde\lambda-\frac{1}{12} \left(R_{ab}K^aK^b 
    \right)_o\, \tilde\lambda^3- \frac{1}{24}  \left(K^aK^bK^c\nabla_c R_{ab}\right)_o\,\tilde\lambda^4+\mathcal{O}\big(\tilde\lambda^5\big).
\label{dAlambda3}
\end{equation}
From now on, we drop the tilde on the affine parameter for convenience, i.e. $\tilde\lambda \to \lambda$.
One can Taylor expand the redshift in terms of the null affine parameter by making use of the relation $1+z=p^a u_a/(p^b u_b)_o = K^a u_a$. Defining the expansion tensor of the matter $\Theta_{ab} = \nabla_b u_a$, we  have:
\begin{equation}
    z=\Xbba_o\lambda
    +\frac{1}{2}\Xbbb_o\lambda^2
    +\frac{1}{6}\Xbbc_o\lambda^3
    +\frac{1}{24}\Xbbd_o\lambda^4
    +\mathcal{O}(\lambda^5),
    \label{z(d)}
\end{equation}
where 
\begin{align}
    \Xbba&\circeq K^a K^b \Theta_{ab}\,,
\label{X1}\\
    \Xbbb&\circeq K^a K^b K^c \nabla_a\Theta_{bc}\,,
\label{X2}\\
    \Xbbc&\circeq K^a K^b K^c K^d \nabla_a\nabla_b\Theta_{cd}\,,
\label{X3}\\
\Xbbd&\circeq K^a K^b K^c K^d K^e \nabla_a\nabla_b\nabla_c\Theta_{de} \,.\label{X4}
\end{align}
We also define
\begin{align}\label{y12}
  \mathbb{Y}^{(1)} = K^a K^b R_{ab}\,, \qquad
  \mathbb{Y}^{(2)} = K^a K^b K^c \nabla_a R_{bc}\,.
\end{align}
Here  the expansion tensor of the matter, 
\begin{equation}
\Theta_{ab} = \frac{1}{3}\Theta h_{ab} + \sigma_{ab}\,,
\end{equation}
describes the volume expansion rate $\Theta$ and the shear rate $\sigma_{ab}$ of a geodesic dust fluid representing the matter flow.
For the interested reader, a deeper explanation of the formalism presented here can be found in \cite{Maartens:2023tib}, which also gives the multipole expansions of $\mathbb{X}^{(1,2)}$  in terms of the matter kinematic parameters. The multipole expansions of $\mathbb{X}^{(3,4)}$ and $\mathbb{Y}^{(2)}$ are presented in Appendix \ref{S:Appendix}.

Proceeding with the distance-redshift relation derivation, relation \eqref{z(d)} can be inverted by using the Lagrange inversion theorem, which gives
\begin{align} \label{lambdaz1}
    \lambda &=\frac{1}{\Xbba_o}\, z
    -\frac{\Xbbb_o}{2\left(\Xbba_o\right)^3} \, z^2
    +\frac{\Big[3\left(\Xbbb_o\right)^2-\Xbba_{o} \Xbbc_{o}\Big]}{6\left(\Xbba_o\right)^5}\, z^3
    \\ \notag &{}\quad
    -\frac{\Big[15\left(\Xbbb_o\right)^3-10\Xbba_{o}\Xbbb_{o}\Xbbc_{o}+\left(\Xbba_o\right)^2\Xbbd_{o}\Big]}{24\left(\Xbba_o\right)^7}\,  z^4+\mathcal{O}(z^5).    
\end{align}
By combining eq. (\ref{lambdaz1}) with eq. (\ref{dAlambda2}), we can find the relation between the angular diameter distance and the redshift:
\begin{align} \label{lambdaz3}
    d_A&=\frac{1}{\Xbba_o}\, z
    -\frac{\Xbbb_o}{2\left(\Xbba_o\right)^3}\,  z^2
    +\frac{\Big[6\left(\Xbbb_o\right)^2-2\Xbba_{o}\Xbbc_{o}-\mathbb{Y}^{(1)}_o\Big]}{12\left(\Xbba_o\right)^5}\, z^3
    \\ \notag
 &{}\quad -\frac{\Big[15\left(\Xbbb_o\right)^3-10\Xbba_{o}\Xbbb_{o}\Xbbc_{o}+\left(\Xbba_o\right)^2\Xbbd_{o}-3\mathbb{Y}^{(1)}_o \left(\Xbba_o\right)^2 \Xbbb_{o}+2\left(\Xbba_o\right)^3 \mathbb{Y}^{(2)}_o\Big]}{24\left(\Xbba_o\right)^7} \, z^4+\mathcal{O}(z^5),
\end{align}
as well as its inverse:
\begin{align}
    z&=\Xbba_o\, d_A+\frac{1}{2}\Xbbb_o\, d_A^2+\frac{1}{12}\left(2\Xbbc_o+\mathbb{Y}^{(1)}_o\Xbba_o\right)d_A^3+ \frac{1}{24}\left(2\mathbb{Y}^{(1)}_o \Xbbb_o-\mathbb{Y}^{(2)}_o \Xbba_o+\Xbbd_o\right)d_A^4\;.
    \label{zfd2}
\end{align}
Rewriting in the form of eq. (\ref{daz2}), we have
\begin{align}
    d_A(z, \mathbf{n})&=\frac{z}{\Hbb_o}\bigg[1-\frac{1}{2}\left(3+\Qbb_o\right)z+\frac{1}{6}\left(11+7\Qbb_o+3\Qbb_o^2+\Rbb-\Jbb_o\right)z^2
    \\ \notag 
    &{}\qquad\quad+\frac{1}{24}\left(-50-10\Rbb-\Qbb_o\big(46+6\Rbb+39 \Qbb_o+15 \Qbb_o^2\big)+\Jbb_o\big(13+10\Qbb_o\big)+\Sbb_o\right)z^3 \bigg],
\end{align}
for the angular diameter distance and 
\begin{align}\label{dL_expf}
    d_L(z, \mathbf{n})&=\frac{z}{\Hbb_o}\bigg[1+\frac{1}{2}\left(1-\Qbb_o\right)z +\frac{1}{6}\left(3\Qbb_o^2+\Qbb_o-1+\Rbb_o-\Jbb_o\right)z^2
    \\ \nonumber
    &{}\qquad\quad+\frac{1}{24}\left(2-\Qbb_o\big(2+15\Qbb_o+15\Qbb_o^2\big)+5\Jbb_o(1+2\Qbb_o) -2\Rbb_o(1+3\Qbb_o)+\Sbb_o\right)z^3\bigg]+\mathcal{O}(z^5) \,,
\end{align}
for the luminosity distance. Here
\begin{equation}
  \mathbb{H}\circeq K^a K^b \Theta_{ab} = \Xbba\,,
\label{Heff10}
\end{equation}

\begin{equation}
  \mathbb{Q}\circeq-3+\frac{K^a K^b K^c \nabla_c\Theta_{ab}}{\mathbb{H}^2} 
   = -3 +\frac{\Xbbb}{\mathbb{H}^2}\,,
\label{Qeff10}
\end{equation}

\begin{equation}
  \mathbb{R} \circeq 1+\mathbb{Q}-\frac{K^a K^b R_{ab}}{2\mathbb{H}^2}  = -2 +\frac{\Xbbb}{\mathbb{H}^2} -\frac{\mathbb{Y}^{(1)}}{2\mathbb{H}^2}\,,
\label{Reff10}
\end{equation}
\begin{equation}
\mathbb{J} \circeq -10\mathbb{Q}-15+\frac{K^a K^b K^c K^d \nabla_c\nabla_d\Theta_{ab}}{\mathbb{H}^3} 
 = 15 -10 \frac{\Xbbb}{\mathbb{H}^2} +\frac{\Xbbc}{\mathbb{H}^3}\,,
\label{Jeff10}
\end{equation}
represent the covariant cosmographic Hubble, deceleration, curvature and jerk parameters
(see also \cite{Sachs:1961zz,Clarkson_theses_2000,heinesen_2021}). Additionally, we introduced a new covariant `snap' parameter: 
\begin{align}
\label{snap_gen}\notag
\Sbb &\circeq 113+17\Jbb+115\Qbb+10\Qbb^2-8\Rbb-\frac{K^a K^b K^c \nabla_c R_{ab}}{\Hbb^2}-\frac{K^c K^d K^e K^a K^b \nabla_c\nabla_d\nabla_e\Theta_{ab}}{\Hbb^4}
\\
&= 129 -\frac{1}{\mathbb{H}^2}\left[ 123\;\Xbbb + \mathbb{Y}^{(2)}\right] + 17\;\frac{\Xbbc}{\mathbb{H}^3} + \frac{1}{\mathbb{H}^4} \left[10 \;\big(\Xbbb\big)^2 -\Xbbd \right]\;.
\end{align}
The multipoles of the first four covariant cosmographic parameters given in equations \eqref{Heff10}--\eqref{Jeff10} are known (see \cite{Kristian:1965sz,Clarkson_theses_2000,heinesen_2021,Mac_Ellis:1970,Maartens:2023tib}). The covariant form of the snap parameter \eqref{snap_gen} is newly presented here. The snap has independent multipoles from $\ell=0$ to $\ell=5$. 
Furthermore, the multipoles of $\Xbbc$ and $\Xbbd$ are given for the first time in Appendix \ref{S:Appendix} in terms of the matter kinematic parameters $\Theta$ and $\sigma_{ab}$. The multipoles of $\mathbb{Y}^{(1,2)}$ are also given in Appendix \ref{S:Appendix}.

\section{Degrees of freedom}\label{sec_dof}

Expanding the distance-redshift relation to fourth order introduces additional degrees of freedom (d.o.f.). 
Each independent multipole contributes a total of $2\ell+1$ d.o.f.
Table \ref{tab_dof} summarizes the d.o.f. for each relevant cosmographic parameter under various assumptions. In the general case, without assuming any symmetry,  
86 fitting parameters are required to fully reconstruct the functional forms of $\mathbb{H}$, $\mathbb{Q}$, $\mathbb{J-R}$ and $\mathbb{S}$ for the case of a non-geodesic fluid. Note that the jerk adds 25 d.o.f. and not 36 as given in \cite{heinesen_2021}. The discrepancy comes from an assumption of the author regarding the multipole $\ell=5$ of the jerk as independent. Here we find that this is actually not the case, since it can be determined by the lower order coefficients (see eq. \eqref{j5eq}).

Motivated by the results of \cite{kalbouneh_marinoni_bel_2023}, which found observational evidence for  the fluctuations of the local universe displaying an axially symmetric configuration, we now determine how this condition reduces the d.o.f. of the covariant cosmographic parameters.

With the axial symmetry assumption, it is enough to  perform a Legendre expansion of physical quantities  in terms of the polar angle $\theta$ alone, instead of the full spherical harmonic expansion \cite{Kalbouneh:2024}:
\begin{equation}
    f=\sum_\ell f_\ell \;P_\ell(\cos\theta).
\end{equation}
Therefore, in this case,  each multipole contributes  one degree of freedom only. 
If one adds the constraint that the motion of the cosmic fluid elements  is geodesic (no 4-acceleration),  the total number of degrees of freedom decreases to 17 (see the third column in Table \ref{tab_dof}). Note that the direction of the axis of symmetry adds 2 degrees of freedom to the total number.

Although the parameters $\Xbb^{(1,2,3,4)}$ have a limited number of multipoles, as shown in appendix \ref{app_X3},  the covariant cosmographic parameters display more multipoles because of the power of $\Hbb$ appearing in the denominator of the defining equations.  These higher multipoles are not new d.o.f. since they can be related to the lower multipoles and they are subdominant -- because they are suppressed by a factor proportional to the ratio of the quadrupole to the monopole of $\Hbb$ \cite{Kalbouneh:2024}, which is much smaller than 1. Here we present their forms in the case of axial symmetry, after linearizing with respect to $\Hbb_2/\Hbb_0$. For $\Qbb$:
\begin{align}
    \Qbb_4 &=-\frac{36\,\Hbb_2}{35\, \Hbb_0}\,\Qbb_2 \,,
\\
    \Qbb_5 &=-\frac{20\,\Hbb_2}{21 \,\Hbb_0}\,\Qbb_3\,,
\end{align}
and for $\ell>5$ the multipoles are second order in $\Hbb_2/\Hbb_0$. 
For the curvature we find 
\begin{align}
    \Rbb_3&=\Qbb_3+\frac{6\,\Hbb_2}{5\, \Hbb_0}(\Qbb_1-\Rbb_1)\,,
\\    \Rbb_4&=-\frac{36\,\Hbb_2}{35\, \Hbb_0}\,\Rbb_2 \,,
\\    \Rbb_5&=-\frac{20\,\Hbb_2}{21 \,\Hbb_0}\,\Qbb_3\,.
\end{align}
For $\ell>5$ the multipoles are second order in $\Hbb_2/\Hbb_0$. For the jerk,  
\begin{align}\label{j5eq}
    \Jbb_5 &=-\frac{10\,\Hbb_2}{21\, \Hbb_0}(3\Jbb_3+10\Qbb_3)\,,
\\    
\Jbb_6 &=-\frac{15\,\Hbb_2}{11\, \Hbb_0}\,\Jbb_4 \,.
\end{align}
For $\ell>6$ the multipoles are second order in $\Hbb_2/\Hbb_0$. For the snap, the maximum independent multipole is $\ell=5$, and the higher multipoles can be expressed as
\begin{align}
    \Sbb_6 &=\frac{1000 \, \Qbb_3^2}{231}+\frac{5 \Hbb_2}{7623\, \Hbb_0} \left(11781 \Jbb_4+15840 \Qbb_1 \Qbb_3+320 \Qbb_3^2-2772 \Sbb_4\right) \,,
\\    \Sbb_7 &=-\frac{252\,\Hbb_2}{143 \,\Hbb_0}\,\Sbb_5 \,,
\\    \Sbb_8 &=-\frac{3200\,\Hbb_2}{429 \,\Hbb_0}\,\Qbb_3^2 \,.
\end{align}
For $\ell>8$ the multipoles are second order in $\Hbb_2/\Hbb_0$.

According to \cite{Heinesen_Macpherson_2022}, the degrees of freedom can be further reduced since only a subset of the multipoles of the covariant cosmographic parameters significantly contribute to the signal.  For each covariant cosmographic parameter these {\it  dominant} multipoles are the monopole and the $\ell \ge 1$ multipoles containing the maximum number of spatial derivatives in each expansion term. Indeed these gradients boost the amplitude proportionally to $\xi_H^{-1} \equiv R_H/R_s \gg 1$, the ratio between the Hubble horizon scale and the size of the fluctuations in the Hubble diagram  \cite{Kalbouneh:2024}.
The dominant multipoles  are shown in the fourth column of Table \ref{tab_dof}.

\begin{table}[h]
\begin{tabular}{|c|c|c|c|}
\hline
\begin{tabular}[c]{@{}c@{}}Covariant cosmographic\\ parameters\end{tabular} & \begin{tabular}[c]{@{}c@{}}General case\\ (including 4-acceleration)\\ d.o.f\end{tabular} & \begin{tabular}[c]{@{}c@{}}With axial symmetry\\ (and geodesic motion)\\ d.o.f.\end{tabular} & \begin{tabular}[c]{@{}c@{}} Dominant multipoles\\ (approximation)\\ d.o.f\end{tabular} \\ \hline\hline
$\Hbb$                                                                     & 9                                                                                         & 2                                                                                             & 2 $(\ell=0,$ $2)$                                                                     \\ \hline
$\Qbb$                                                                     & 16                                                                                        & 4                                                                                             & 3 ($\ell=0$, $1$, $3$)                                                              \\ \hline
$\Jbb-\Rbb$                                                                & 25                                                                                        & 5                                                                                             & 3 ($\ell=0$, $2$, $4$)                                                                \\ \hline
$\Sbb$                                                                     & 36                                                                                        & 6                                                                                             & 4 ($\ell=0$, $1$, $3$, $5$)                                                         \\ \hline\hline
Total                                                                      & 86                                                                                        & 17                                                                                            & 12                                                                                    \\ \hline
\end{tabular}
\caption{Degrees of freedom associated with specific covariant cosmographic parameters. The maximum independent multipole is $\ell=2$, $3$, $4$, $5$ for $\mathbb{H}$, $\mathbb{Q}$, $\mathbb{J-R}$ and $\mathbb{S}$ respectively.}
\label{tab_dof}
\end{table}

\section{Numerical example for an analytical model}\label{sec_num}

The expansion rate fluctuation field $\eta$ \cite{kalbouneh_marinoni_bel_2023,Kalbouneh:2024}
is an observable specifically tailored to identify and classify 
angular anisotropies in the Hubble diagram and to constrain the value of covariant cosmographic parameters.  
It is straightforward to expand $\eta$ in a power series of the redshift up to  $O(z^4)$ and predict how its amplitude depends on the snap parameter.  We find that
\begin{align}
  \eta (z,\boldsymbol{n}) &=\log \mathbb{H}(\boldsymbol{n})-\mathcal{M}(z) -\frac{1-\mathbb{Q}(\boldsymbol{n})}{2 \ln 10}\, z
  +\frac{7-\mathbb{Q}(\boldsymbol{n}) \big[10+9\mathbb{Q}(\boldsymbol{n})\big] +4\big[\mathbb{J}(\boldsymbol{n})-\mathbb{R}(\boldsymbol{n})\big]}{24\ln 10}\, z^2 
  \\ \notag
  &~~~~ +\frac{1}{24 \ln10}\bigg\{-5 \Jbb (\boldsymbol{n}) \big[2 \Qbb(\boldsymbol{n})+1\big]+2 \big[\Jbb(\boldsymbol{n})-\Rbb(\boldsymbol{n})\big] \big[\Qbb(\boldsymbol{n})-1\big]
  \\ \notag
  &\qquad\qquad\qquad +\Qbb(\boldsymbol{n}) \Big[9+2 \Qbb(\boldsymbol{n}) \big[5 \Qbb(\boldsymbol{n})+8\big]
  +6 \Rbb(\boldsymbol{n})\Big]+2 \Rbb(\boldsymbol{n})-5-\Sbb(\boldsymbol{n})\bigg\}\,z^3+
   \mathcal{O}(z^4)\,. 
\label{eta_exp_1}
\end{align}

In this section, we calculate the amplitude of the snap parameter for the $M2$ model of perturbation fluctuations in the distance-redshift relation introduced by \cite{Kalbouneh:2024}. 
This is an analytical and realistic model of the local universe, embodying axial symmetry and accurately describing observations. We use it to assess whether the snap contribution to the expansion rate fluctuations  $\eta$  can be observationally constrained by future surveys, such as ZTF  \cite{amenouche:tel-04165406}.

\begin{figure*}
  \centering
  \includegraphics[scale=0.6]{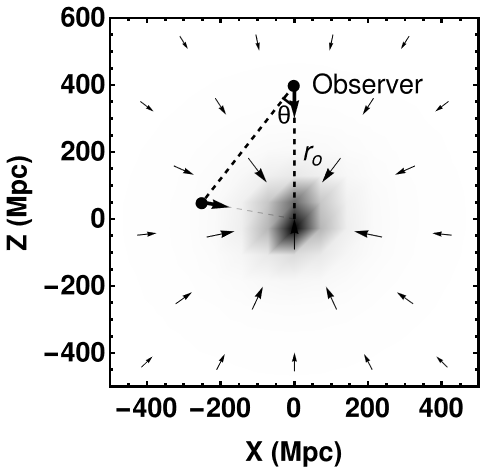}
  \caption{The analytical model $M2$ with a spherical density contrast projected onto a plane. We  show the observer's position relative to the attractor (which is at the center) and the peculiar velocity perturbation caused by the attractor. The angle $\theta$ and radial position $r_o$ are also indicated.}
  \label{modexp}
\end{figure*}

The $M2$  model describes an Einstein-de Sitter  spacetime perturbed by a spherically symmetric density contrast
\begin{equation}
    \delta(r,t_0) = \delta_c \left[ 1 + \left(\frac{r}{R_s}\right)^2 \right]^{-3/2},
\label{delta_0}
\end{equation}
where $\delta_c$ is the density contrast at the center of the perturbation and $R_s$ is the typical radial extension of the perturbation. The observer is off-center at  $r_o$, and sees an axially symmetric universe. The $M2$ model is defined by (see Figure \ref{modexp})
\begin{equation}\label{m2mod}
    M2: \quad r_o = 400 \,\text{Mpc}, \quad \delta_c = 2.5, \quad R_s = 56\,{\rm Mpc},
\end{equation}
with $H_0=70$\,km/s/Mpc in the background. As discussed before, axial symmetry allows us to perform a Legendre expansion in terms of the polar angle $\theta$ instead of the full spherical harmonic expansion \cite{Kalbouneh:2024}.

\begin{figure*}
  \centering
  \includegraphics[scale=0.43]{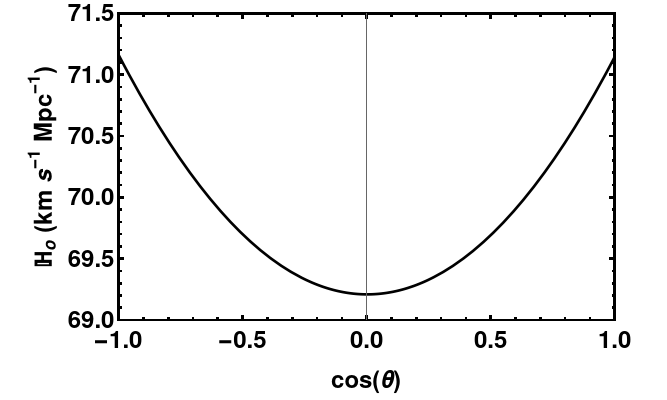}
  \includegraphics[scale=0.43]{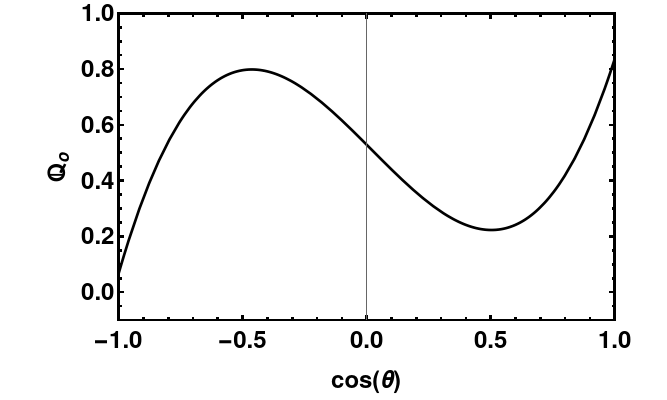}
  \includegraphics[scale=0.43]{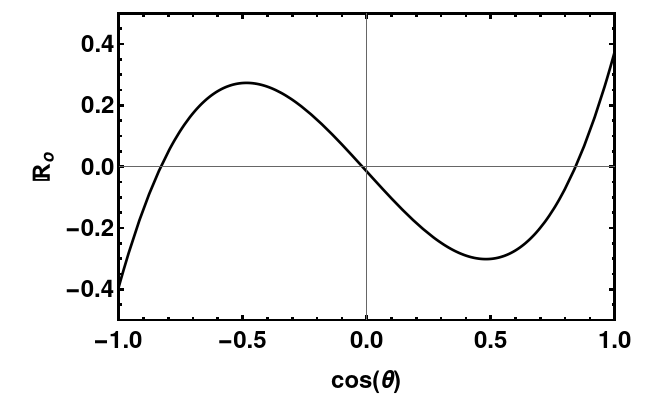}
  \includegraphics[scale=0.43]{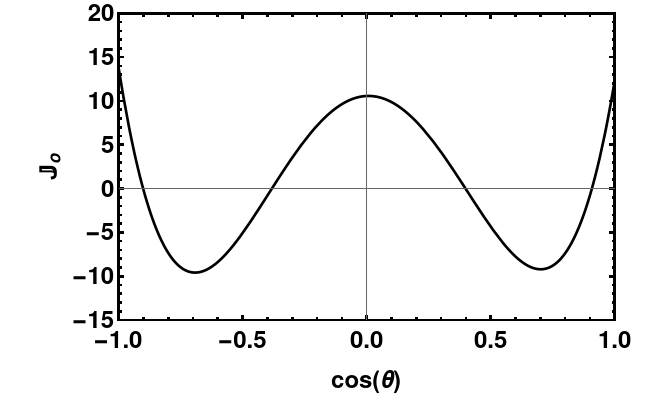}
  \includegraphics[scale=0.43]{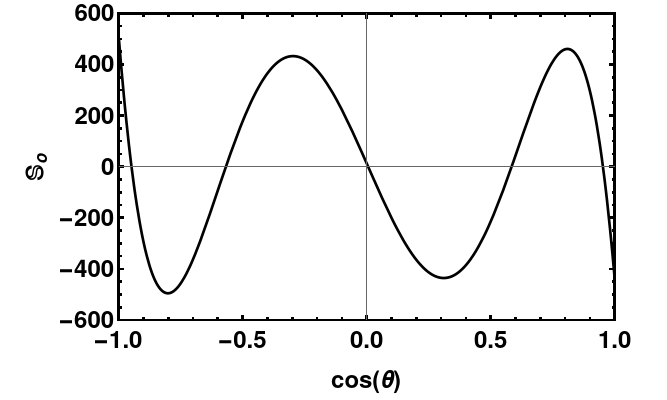}
  \caption{Cosmographic parameters for the analytical model $M2$ [see \eqref{m2mod}], as a function of  $\cos\theta$.}
  \label{HABC_2}
\end{figure*}

The multipoles of the cosmographic parameters for this model are given in \cite{Kalbouneh:2024} without the snap. In Figure \ref{HABC_2}, we show the cosmographic parameters including the snap for the model as a function of $\cos\theta$, computed by using the equations \eqref{Heff10}--\eqref{snap_gen}, and the metric
\begin{equation}
\mathrm{d}s^2=-\big(1+2\Phi\big)\mathrm{d}t^2+a^2(t)\big(1-2\Phi\big)\left[\mathrm{d}r^2+ r^2(\mathrm{d}\psi^2+\sin^2\psi \,\mathrm{d} \phi^2)\right]\;.
\label{metricper}
\end{equation}
Here
\begin{equation}
\Phi(r)=\Phi_c \;\frac{R_s}{r}\sinh^{-1}\left(\frac{r}{R_s}\right) \quad \mbox{with} \quad \Phi_c=-\frac{3}{2}H^2_0 R_s^2 \delta_c\,,
\label{phi_r}
\end{equation}
is the exact solution  in a Newtonian approximation for the Poisson equation with the density contrast given by \eqref{delta_0}, and assuming that only the growing mode is of interest. The 4-velocity $u^\mu$ is
\begin{equation}
u^\mu=\left(1-\Phi, \frac{v}{a},0,0 \right)\quad \text{where}\quad v(r,t)=-\frac{2}{3 a(t) H(t)}\frac{\partial \Phi(r,t)}{\partial r} \,.
\end{equation}
Here $(v/a)\,\delta^\mu_r$ is the peculiar velocity field \cite{Pee1980}. 
The multipoles of the snap $\Sbb_\ell$ at linear order for this model are 
\begingroup
\allowdisplaybreaks
\begin{align}
    \Sbb_0 &=-\frac{7}{2}-\frac{\delta _c \left[200 \xi _H^2 \left(\xi _o^2+1\right){}^2 \xi _o^3-81 \xi _o^7+54 \xi _o^5\right]}{30 \xi _H^2 \left(\xi _o^2+1\right){}^{7/2}},
\label{S0f}\\
    \Sbb_1 &=-\frac{9 \delta _c \xi _o^4}{35 \xi _H^3 \left(\xi _o^2+1\right){}^{9/2}}\left[63 \xi _H^2 \left(\xi _o^2+1\right){}^2+25 \left(5 \xi _o^2-2\right) \xi _o^2\right],
\label{S1f}\\
    \Sbb_2 &=\frac{10}{21} \delta _c \xi _o^3 \left[\frac{-27 \xi _o^2/\xi _H^2-28 \left(\xi _o^2+1\right){}^2 \left(3 \xi _o^2+4\right)}{\left(\xi _o^2+1\right){}^{7/2}}+84\,
   \text{csch}^{-1}\left(\xi _o\right)\right],
\label{S2f}\\
    \Sbb_3 &=\frac{4 \delta _c \xi _o^4}{15 \xi _H^3} \left[\frac{175 \xi _o^2-6 \xi _H^2 \left(\xi _o^2+1\right){}^2 \left(15 \xi _o^4+35 \xi _o^2+23\right)}{\left(\xi _o^2+1\right){}^{9/2}}+90\, \xi
   _H^2\, \text{csch}^{-1}\left(\xi _o\right)\right],
\label{S3f}\\
    \Sbb_4 &=\frac{12 \delta _c \xi _o^5}{35 \xi _H^2} \left[\frac{-7 \left(15 \xi _o^4+50 \xi _o^2+58\right) \xi _o^2-176}{\left(\xi _o^2+1\right){}^{7/2}}+105 \,\text{csch}^{-1}\left(\xi
   _o\right)\right],
\label{S4f}\\
    \Sbb_5 &=\frac{8 \delta _c \xi _o^6}{21 \xi _H^3}\left[\frac{3 \left(105 \xi _o^6+455 \xi _o^4+756 \xi _o^2+582\right) \xi _o^2+563}{\left(\xi _o^2+1\right){}^{9/2}}-315\, \text{csch}^{-1}\left(\xi
   _o\right)\right].
\label{S5f}
\end{align}
\endgroup
Here $\xi_o=R_s/r_o$ and $\xi_H=R_s/R_H=H_0 R_s$ are dimensionless parameters, and $\xi_H$ is the ratio between the scale of the perturbation over the Hubble horizon as defined in section \ref{sec_dof}. For our model, the true values for the multipoles of the snap are ($-4.8$, $93$, $-9$, $363$, $29$, $-936$) for $\ell=0$ to $\ell=5$.
The multipoles $\ell=1$, $3$ and $5$  are dominant relative to the others since they have higher-order spatial derivatives as discussed in section \ref{sec_dof}. This clearly appears in eqs. \eqref{S0f}--\eqref{S5f}, where the dominant multipoles have $\xi_H^3$ in the denominator but not the others.

The relationship between the multipoles of the expansion rate fluctuation field $\eta$ and the multipoles of the covariant cosmographic parameters is detailed in Appendix \ref{app_eta_Xl}. We assess how well these theoretical predictions describe simulated data using a $\chi^2$ statistical analysis, following the procedure detailed in Paper II.
To this end, we randomly sample the analytical $M2$ model to construct a mock catalog of $N=30,000$ redshift and distance measurements for objects isotropically distributed across the sky. The redshift distribution is generated to match the expected distribution from the ZTF survey~\cite{amenouche:tel-04165406} in the range $0.01<z<0.1$, and the error in the distance modulus is assumed to be 0.15 \cite{Dhawan_2022} with no correlation between the measurements. The redshift and the luminosity distance in the model are computed using eqs. (5.11) and (5.13) of \cite{Kalbouneh:2024} (see also \cite{Hui:2005nm}).

\begin{figure*}
  \centering
  \includegraphics[scale=0.3]{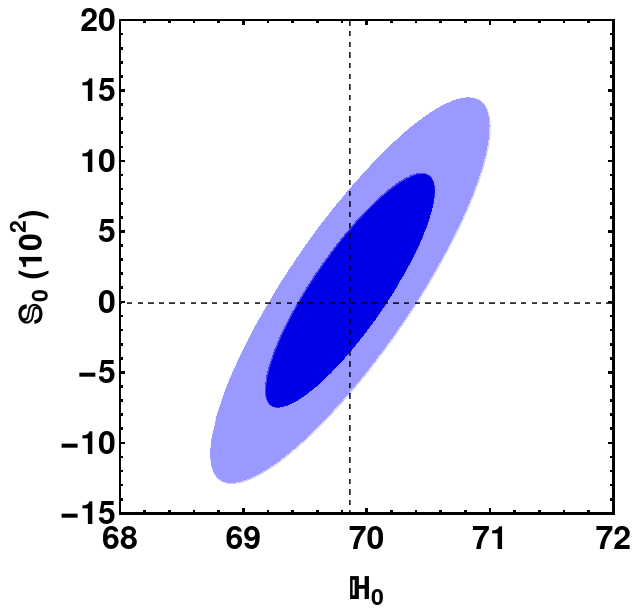}
  \includegraphics[scale=0.3]{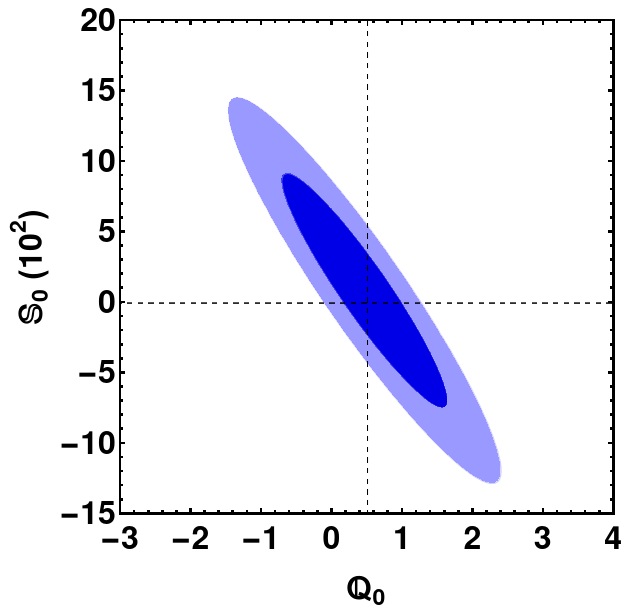}
  \includegraphics[scale=0.3]{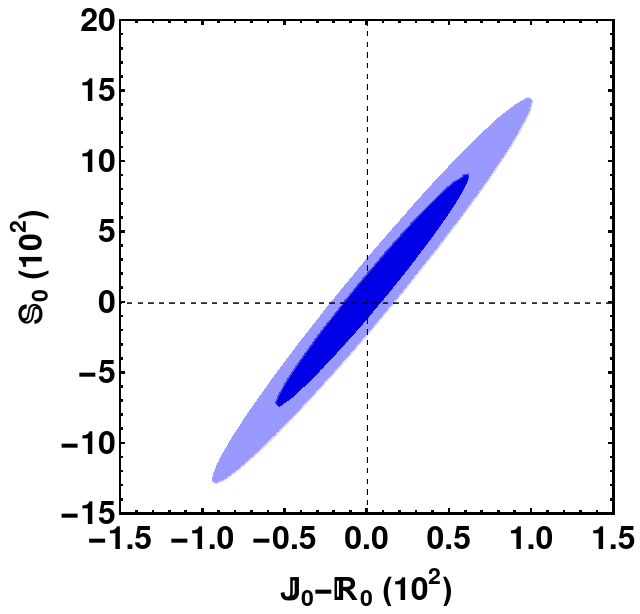}
\includegraphics[scale=0.3]{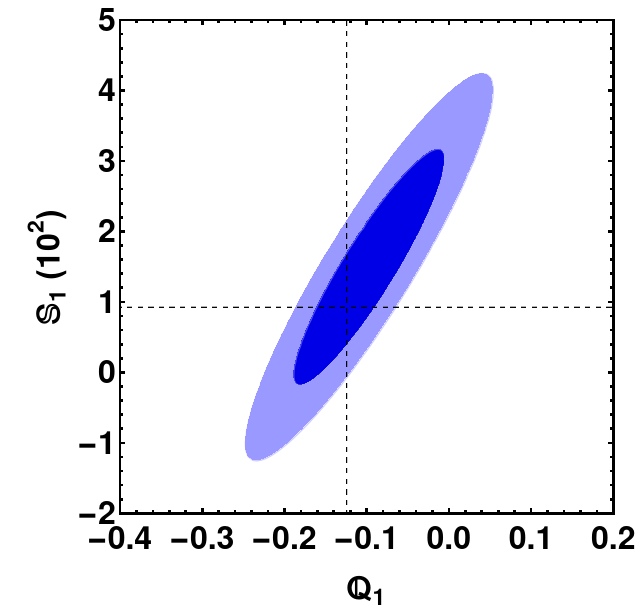}
    \includegraphics[scale=0.3]{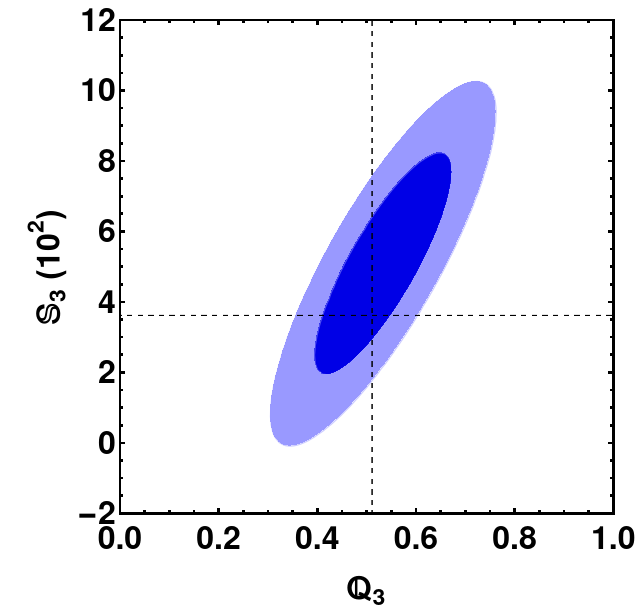}
 \includegraphics[scale=0.3]{fig5.png}
  \caption{Likelihood of the multipoles of the snap as reconstructed for the analytical model $M2$. The intersection of the dashed lines presents the true input value. The subscript on the cosmographic parameters indicates the multipole $\ell$.}
  \label{likS}
\end{figure*}

Figure \ref{likS} shows the likelihood of the reconstructed monopole and dominant multipoles of the snap for model $M$. 
It is worthy to point out that, due to the well know degeneracy problem in the cosmographic approach~\cite{Shafieloo2011} between the jerk, snap and higher order parameters with the curvature term, we have ignored the $\mathbb{R}_0$ term in equation \eqref{monom}. This choice also is not expected to impact the final results given that $\mathbb{R}_0$ is not a dominant term.
By adding the snap correction, the estimation of the dominant multipoles up to $\ell=3$ is unbiased, although this comes at the cost of larger error contours in the recovered parameters. As a rule of thumb, should one consider a survey with a different number $N$,  the errors must be re-scaled by a factor of $\sim\sqrt{30,000/N}$. 

The other parameters for the same analytical model are computed in \cite{Kalbouneh:2024}, where it is shown that $\Qbb_3$ and $\Jbb_4$ are biased. Here, after adding the snap, $\Qbb_3$ is no longer biased, but $\Jbb_4$ remains unaffected because $\Sbb_4$ is not dominant and can be neglected. Therefore, to avoid the bias in $\Jbb_4$, one needs to add the crackle, i.e.the fifth-order covariant cosmographic parameter. However, Figure \ref{likS} also shows that the recovered amplitude of the multipole $\Sbb_5 = -170 \pm 68$  is significantly different from the input value ($-936$).  In the same spirit advocated in this paper, to correct for this residual discrepancy,  a  higher-order expansion, to $\mathcal{O}(z^6)$, of the distance as a function of redshift is needed.  The inclusion of the pop (sixth-order covariant cosmographic parameter)  would help unbias  the snap multipole $\ell=5$.

The amplitude and size of structures in the local universe constrain the redshift range in which the cosmographic expansion can be safely applied. However, in our actual universe, and without any prior knowledge about its structure, this range can be constrained using two strategies, as discussed in \cite{Kalbouneh:2024} in Section 8. First, we conduct a $\chi^2$-test to determine the redshift interval within which all observed multipoles of the expansion rate fluctuation field $\eta$, up to a specified expansion order in redshift, align consistently with theoretical predictions. Second, we verify that adjusting the lower or upper bounds within this redshift range does not substantially impact the estimated values of the best-fitting cosmographic parameters. These steps collectively ensure the robustness and accuracy of the recovered parameter values.

\section{Conclusion}

The recent challenges  faced by the standard model of cosmology, together with an enormous increase of available data,  create a pressing need to test the foundations of the standard model of cosmology, in particular, the Cosmological Principle. In this paper, we extend the model-independent covariant cosmographic approach developed in papers I  and II \cite{Maartens:2023tib,Kalbouneh:2024} (see also \cite{Kristian:1965sz, MacCallum_Ellis_1970, Clarkson_theses_2000,clarkson_maartens_2010,Umeh:2013, heinesen_2021}), by including the fourth-order term of the covariant luminosity distance expansion in redshift, \eqref{dL_expf}.
The motivation to do so arises from the poor estimation of the dipole $\mathbb{Q}_1$ and octupole $\mathbb{Q}_3$ of the deceleration parameter, even when the third order expansion $\mathbb{J}$ term is taken into account, as shown in \cite{Kalbouneh:2024}. 
The fourth-order expansion term has a new cosmographic parameter, the snap $\Sbb$. 
Its relation to the observed lightray direction, the matter fluid 4-velocity and the curvature of the space is given in eq. (\ref{snap_gen}). 

In Section \ref{sec_dof}, we present the number of degrees of freedom in this cosmographic approach. The new snap term includes extra 36 d.o.f. in the case of a generic spacetime. We show, however, that this number can be significantly reduced  by considering only the  dominant multipoles  (see also \cite{Heinesen_Macpherson_2022}) and by assuming  axial symmetry. The results are presented in Table \ref{tab_dof}.

In Section \ref{sec_num} we perform a forecast for the precision in measuring the multipoles of the snap, using an analytical model motivated by observational evidence.
The likelihoods for the snap multipoles are presented in Figure \ref{likS}. The estimation of all the multipoles up to $\ell =3$ are unbiased.
However, the estimation of the $\ell = 5$ multipole is biased, which likely requires a higher-order expansion in the $d_L(z,\mathbf{n})$ series in order to be corrected.

Finally, returning to the initial motivation, we have verified our initial prediction, showing that the inclusion of the snap term does indeed improve the likelihood of both the dipole $\mathbb{Q}_1$ and octupole $\mathbb{Q}_3$ of the deceleration parameter (see Figure \ref{HABC_1}). Furthermore, the reason for this is now clearer. As discussed in Section \ref{sec_num} and shown in Table \ref{tab_dof}, the dominant multipoles of the snap $\mathbb{S}$ are precisely  $\ell =$ 0, 1, 3 and 5, in accordance with $\ell =$ 0, 1 and 3 for the case of the deceleration  $\mathbb{Q}$. 
On the other hand, the dominant multipoles for the jerk $\mathbb{J}$ are $\ell =$ 0, 2 and 4 -- explaining why the inclusion of these terms had only a minor effect on reducing the bias in $\mathbb{Q}_{1,3}$. Following this same logic, one  may expect that, in order to improve the bias found in the multipole $\mathbb{S}_5$, including the fifth-order cosmographic parameter (the crackle) should produce very little effect -- requiring instead the sixth-order term (the pop) in order to correct it. Given the finite amount of data, this will be a task for the future.

\vfill

\acknowledgments 
BK, CM and MS are supported by the {\it Programme National Cosmologie et Galaxies} (PNCG) and {\it Programme National Gravitation Références Astronomie Métrologie} (PNGRAM), of CNRS/INSU with INP and IN2P3, co-funded by CEA and CNES. They also received support from the {\it Agence Nationale de la Recherche} under the grant ANR-24-CE31-6963-01, and the French government under the France 2030 investment plan, as part of the Initiative d’Excellence d'Aix-Marseille Université -  A*MIDEX (AMX-19-IET-012). JS is supported by the Taiwan National Science \& Technology Council.
RM is supported by the South African Radio Astronomy Observatory and the National Research Foundation (grant no. 75415).

\newpage

\appendix

\section{Multipole expansions of $\mathbb{X}^{(3,4)}$ and $\mathbb{Y}^{(1,2)}$}\label{S:Appendix}

$\mathbb{X}^{(1-4)}$ are defined in eqs. \eqref{X1}--\eqref{X4} and $\mathbb{Y}^{(1,2)}$ in eq. \eqref{y12}. Their multipoles are needed to obtain the multipole expansion of the covariant snap parameter.
\vspace*{0.2cm}

\noindent{\em Notation:} Angled brackets on indices indicate the projected, symmetric, tracefree (PSTF) part of a tensor, relative to the matter 4-velocity $u^a$. 
For a vector and a
 rank 2 tensor:
\begin{equation}
    W_{\langle a\rangle}=h_a^{\,b}W_b\,,\quad
    W_{\langle ab \rangle} = \Big[h_{a}^{\,(c} h_{b}^{\,d)}-\frac{1}{3} h_{ab}h^{cd} \Big]W_{cd}\,,
\end{equation}
where  $h_{ab} = g_{ab} + u_a u_b$. 
For the PSTF parts  of higher-rank tensors, see \cite{Maartens:2023tib}.
Note that contraction of an arbitrary tensor with a PSTF tensor isolates the PSTF part:
\begin{equation}
   V^a W_{\langle a\rangle} =V^{\langle a\rangle} W_{\langle a\rangle}\,,\quad
   V^{ab} W_{\langle ab \rangle} = 
   V^{\langle ab \rangle} W_{\langle ab \rangle}\,,
\end{equation}
and similarly for higher rank tensors.

The covariant multipoles of $\mathbb{X}^{(1-4)}$ and $\mathbb{Y}^{(1,2)}$ are all PSTF tensors. In order to simplify the expressions, 
 we use the short-hand notation
\begin{align}
W_{a_1\cdots a_n} \peq V_{a_1\cdots a_n}\quad \Leftrightarrow \quad
    W_{\langle a_1\cdots a_n\rangle}= V_{\langle a_1\cdots a_n\rangle}\,.
\end{align}

\subsection*{Multipoles of $\Xbbc$}\label{app_X3}

\vspace{0.3cm}

\begin{equation}
    \Xbbc = \langle\Xbbc\rangle + \Xbbc_a n^a + \Xbbc_{ab}n^{\langle a}n^{b\rangle} + \Xbbc_{abc}n^{\langle a}n^bn^{c\rangle} + \Xbbc_{abcd}n^{\langle a}n^bn^cn^{d\rangle}\;,
\end{equation}
where 
\begin{align}
    \langle\Xbbc\rangle &= \tfrac{8}{5} \sigma_{ab}\sigma^{bc}\sigma^a{}_c + \tfrac{8}{5}  \Theta \sigma_{ab}\sigma^{ab} + \tfrac{2}{9} \Theta ^3 - \tfrac{2}{3} \Theta \dot{\Theta } - \tfrac{12}{5} \sigma ^{ab} \dot\sigma _{ab} 
     + \tfrac{1}{3} \ddot{\Theta } + \tfrac{2}{15} \nabla _{a}\nabla _{b}\sigma ^{ab} 
     \\&\;\;
     + \tfrac{1}{9} \nabla^{a}\nabla _{a}\Theta  + \tfrac{1}{9} u^{a}  u^{b} \nabla _{a}\nabla _{b}\Theta\,, \nonumber
    \\
   \Xbbc_a & \peq \tfrac{6}{5} \sigma ^{bc} \nabla _{a}\sigma _{bc} + \tfrac{2}{3} \Theta  \nabla _{a}\Theta  + \tfrac{4}{5} \Theta  \nabla _{b}\sigma _{a}{}^{b} - \tfrac{2}{3}  u^{b} \nabla _{b}\nabla _{a}\Theta  - \tfrac{2}{5}  u^{b} \nabla _{b}\nabla _{c}\sigma _{a}{}^{c} + \tfrac{8}{15} \sigma _{ab} \nabla ^{b}\Theta  
   \\&\;\;
   + \tfrac{6}{5} \sigma ^{bc} \nabla _{b}\sigma _{ca}
   + \tfrac{6}{5} \sigma _{ab} \nabla _{c}\sigma^{bc} - \tfrac{2}{5}  u^{b} \nabla _{c}\nabla _{b}\sigma _{a}{}^{c}\,,
   \nonumber \\
   \Xbbc_{ab} & \peq \tfrac{32}{7} \sigma _{a}{}^{c} \sigma _{b}{}^{d} \sigma _{cd} + \tfrac{100}{21} \Theta \sigma _{a}{}^{c} \sigma _{b c}  + \tfrac{4}{63}\big(29 \Theta ^2  - 21\dot{\Theta }\big)\sigma_{ab}  + \tfrac{1}{3} \nabla _{a}\nabla _{b}\Theta       - \tfrac{46}{21} \Theta  \dot\sigma _{ ab} 
   \\&\;\;
   - \tfrac{48}{7} \sigma _{ a}{}^{c} \dot \sigma _{b c} + \tfrac{2}{7} \nabla _{c}\nabla _{a}\sigma _{b}{}^{c} + \tfrac{1}{7} \nabla^{c}\nabla _{c}\sigma _{ab} 
   + \tfrac{8}{7}  u^{c} u^{d} \nabla _{c}\nabla _{d}\sigma _{ab} + \tfrac{2}{7} \nabla _{a}\nabla _{c}\sigma _{b}{}^{c}\,,
   \nonumber \\
   \Xbbc_{abc} & \peq \tfrac{4}{3} \Theta\sigma _{a b } \nabla _{c}  + 2 \Theta  \nabla _{a}\sigma _{bc}  + 6 \sigma _{a}{}^{d} \nabla _{b }\sigma _{c d} - u^{d} \nabla _{ a }\nabla _{b}\sigma _{cd} -  u^{d} \nabla _{d}\nabla _{ a }\sigma _{bc}\,,\\
   \Xbbc_{abcd} & \peq \tfrac{2}{3} \sigma _{ ab} \sigma _{cd } \Theta  + \nabla _{a}\nabla _{b}\sigma _{cd}\,.
\end{align}

\vspace*{0.05cm}
\subsection*{Multipoles of $\Xbbd$} \label{app_X4}
\vspace*{0.1cm}
\begin{equation}
    \Xbbd = \langle\Xbbd\rangle + \Xbbd_a n^a + \Xbbd_{ab}n^{\langle a}n^{b\rangle} + \Xbbd_{abc}n^{\langle a}n^bn^{c\rangle} + \Xbbd_{abcd}n^{\langle a}n^bn^cn^{d\rangle} + \Xbbd_{abcde}n^{\langle a}n^bn^cn^dn^{e\rangle}\,,
\end{equation}
where 
\allowdisplaybreaks
\begingroup
\begin{align}
    \langle\Xbbd\rangle &=  
    \tfrac{8}{3} \sigma_{ab} \sigma^{bc} \sigma_{cd} \sigma^{ad} + \tfrac{172}{45} \Theta \sigma_{ab}\sigma^{bc}\sigma^a{}_c   + \tfrac{2}{45}\big(53 \Theta^2- 6 \dot\Theta\big)\sigma_{ab}\sigma^{ab}  +\tfrac{8}{27} \Theta ^4 -  
    \tfrac{34}{27} \Theta ^2 \dot{\Theta } + \tfrac{26}{27} \dot{\Theta }^2 
    \nonumber\\ & \;\;
    - \tfrac{64}{9}\Theta \sigma ^{ab}   \dot\sigma _{ab}
    - \tfrac{52}{5} \sigma _{ab} \sigma ^{bc} \dot\sigma^a{}_{c} 
    + \tfrac{4}{3} \Theta  \ddot{\Theta } - \tfrac{2}{15}  u^{a} \nabla _{a}\nabla _{b}\nabla _{c}\sigma ^{bc} - \tfrac{1}{9} u^{a} \nabla _{a}\nabla^{b}\nabla _{b}\Theta  + \tfrac{8}{27} \nabla _{a}\Theta  \nabla ^{a}\Theta 
    \nonumber\\ & \;\;
     +\tfrac{1}{135}\big(30 \dot{\Theta }-43 \Theta ^2 \big) u^{a} \nabla _{b}\sigma _{a}{}^{b} 
     + \tfrac{2}{3} \nabla_{a}\Theta  \nabla _{b}\sigma^{ab}
    + \tfrac{2}{5} \nabla _{a}\sigma ^{bc} \nabla^{a}\sigma _{bc} + \tfrac{16}{5}  u^{a}  u^{b} \nabla _{a}\sigma ^{cd} \nabla _{b}\sigma _{cd}
    \nonumber\\&\;\;
    + \tfrac{16}{45} \Theta  \nabla _{a}\nabla _{b}\sigma ^{ab} + \tfrac{8}{15} \sigma ^{ab} \nabla^{c}\nabla _{c}\sigma _{ab} + \tfrac{64}{15} \sigma ^{ab}  u^{c}  u^{d} \nabla _{c}\nabla _{d}\sigma _{ab} + \tfrac{32}{45} \sigma ^{ab} \nabla _{b}\nabla _{a}\Theta 
    + \tfrac{4}{9} \Theta  \nabla^{a}\nabla _{a}\Theta
    \nonumber\\&\;\;
    - \tfrac{2}{9} \nabla^{a}\nabla _{a}\dot{\Theta } - \tfrac{2}{3}  u^{a}  u^{b} \nabla _{a}\nabla _{b}\dot{\Theta }
    + \tfrac{8}{15} \sigma ^{ab} \nabla _{b}\nabla _{c}\sigma _{a}{}^{c} + \tfrac{11}{45} \Theta   u^{a}  u^{b} \nabla _{b}\nabla _{c}\sigma _{a}{}^{c} + \tfrac{7}{15} \sigma ^{cd}  u^{a}  u^{b} \nabla _{b}\nabla _{d}\sigma _{ac} 
    \nonumber\\&\;\;
    - \tfrac{49}{45} \Theta u^a \sigma ^{bc}  \nabla _{c}\sigma _{ab} +\tfrac{4}{45}\big(7\dot\Theta- 10\Theta ^2\big)  u^{a}  u^{b}  u^{c} \nabla _{a}\sigma _{bc} 
    + \tfrac{2}{5} \nabla _{a}\sigma ^{ab} \nabla _{c}\sigma _{b}{}^{c} - \tfrac{2}{15}  u^{a} \nabla _{b}\nabla _{a}\nabla _{c}\sigma ^{bc}
    \nonumber\\&\;\;
    + \tfrac{8}{15} \sigma ^{ab} \nabla _{c}\nabla _{b}\sigma _{a}{}^{c} 
    + \tfrac{11}{45} \Theta   u^{a}  u^{b} \nabla _{c}\nabla _{b}\sigma _{a}{}^{c} - \tfrac{2}{15}  u^{a} \nabla _{b}\nabla _{c}\nabla _{a}\sigma ^{bc} + \tfrac{2}{5} \nabla _{a}\sigma _{bc} \nabla ^{c}\sigma ^{ab} + \tfrac{5}{9}  u^{a}  u^{b} \nabla _{b}\sigma _{ac} \nabla ^{c}\Theta
    \nonumber\\&\;\;
    + \tfrac{2}{45}  u^{a}  u^{b} \nabla _{c}\sigma _{ab} \nabla ^{c}\Theta  
    - \tfrac{14}{15} \sigma _{b}{}^{d} \sigma ^{bc}  u^{a} \nabla _{d}\sigma _{ac} + \tfrac{1}{5}  u^{a}  u^{b} \nabla _{c}\sigma _{a}{}^{d} \nabla^c\sigma _{bd} +  u^{a}  u^{b} \nabla _{b}\sigma _{a}{}^{c} \nabla _{d}\sigma _{c}{}^{d}
    \nonumber\\&\;\;
    + \tfrac{14}{5}  u^{a}  u^{b}  u^{c}  u^{d} \nabla _{b}\sigma _{a}{}^{e} \nabla _{d}\sigma _{ce} + \tfrac{7}{15} \sigma ^{cd}  u^{a}  u^{b} \nabla _{d}\nabla _{b}\sigma _{ac} 
     + \tfrac{34}{45} \Theta   u^{a}  u^{b}  u^{c}  u^{d} \nabla _{d}\nabla _{c}\sigma _{ab}
    \nonumber\\&\;\;
    +  u^{a}  u^{b} \nabla _{b}\sigma _{cd} \nabla ^{d}\sigma _{a}{}^{c} + \tfrac{2}{45} \Theta   u^{a}  u^{b} \nabla^{c}\nabla _{c}\sigma _{ab}\,,
   \\ \nonumber\;\;& \\
   \Xbbd_a & \peq 
   \tfrac{344}{105} \Theta  \sigma ^{bc} \nabla _{a}\sigma _{bc} + \tfrac{1}{3} \Theta ^2  u^{b}  u^{c} \nabla _{a}\sigma _{bc} - \tfrac{4}{21} \dot{\Theta }  u^{b}  u^{c} \nabla _{a}\sigma _{bc} + \tfrac{24}{5} \sigma _{b}{}^{d} \sigma ^{bc} \nabla _{a}\sigma _{cd} + \tfrac{122}{105} \sigma_{bc}\sigma^{bc} \nabla _{a}\Theta  + \tfrac{191}{135} \Theta ^2 \nabla _{a}\Theta  
   \nonumber\\ & \;\; \nonumber
   - \tfrac{1}{9} \Theta  \nabla _{a}\dot{\Theta } - \tfrac{64}{35} \sigma ^{cd}  u^{b} \nabla _{a}\nabla _{b}\sigma _{cd} - \tfrac{1}{15} \Theta   u^{b} \nabla _{a}\nabla _{c}\sigma _{b}{}^{c} + \tfrac{2}{35} \nabla _{a}\nabla _{c}\nabla _{b}\sigma ^{bc} + \tfrac{1}{15}\nabla _{a}\nabla _{b}\nabla^b\Theta  - \tfrac{1}{7} \sigma ^{cd}  u^{b} \nabla _{a}\nabla _{d}\sigma _{bc}   
   \\&\;\; \nonumber
   - \tfrac{23}{105} \Theta   u^{b}  u^{c}  u^{d} \nabla _{a}\nabla _{d}\sigma _{bc} + \tfrac{416}{315} \Theta ^2 \nabla _{b}\sigma _{a}{}^{b} - \tfrac{148}{105} \dot{\Theta } \nabla _{b}\sigma _{a}{}^{b} - \tfrac{96}{35}  u^{b} \nabla _{a}\sigma ^{cd} \nabla _{b}\sigma _{cd} - \tfrac{64}{35} \sigma ^{cd}  u^{b} \nabla _{b}\nabla _{a}\sigma _{cd} 
   \\&\;\; \nonumber
   + \tfrac{2}{5}  u^{b} \nabla _{b}\nabla _{a}\dot{\Theta } - \tfrac{124}{105} \Theta   u^{b} \nabla _{b}\nabla _{c}\sigma _{a}{}^{c} - \tfrac{64}{35} \sigma ^{cd}  u^{b} \nabla _{b}\nabla _{d}\sigma _{ac} - \tfrac{64}{35} \sigma _{a}{}^{c}  u^{b} \nabla _{b}\nabla _{d}\sigma _{c}{}^{d} + \tfrac{171}{35} \sigma _{a}{}^{c} \sigma _{bc} \nabla ^{b}\Theta  
   \nonumber\\&\;\;
   + \tfrac{1}{45}  \sigma _{ab} \Theta  \nabla ^{b}\Theta  -2 \sigma _{ab} \nabla ^{b}\dot{\Theta } + \tfrac{356}{105} \sigma ^{bc} \Theta  \nabla _{c}\sigma _{ab} + \tfrac{509}{315} \Theta ^2  u^{b}  u^{c} \nabla _{c}\sigma _{ab} - \tfrac{116}{105} \dot{\Theta }  u^{b}  u^{c} \nabla _{c}\sigma _{ab} + \tfrac{356}{105} \sigma _{a}{}^{b} \Theta  \nabla _{c}\sigma _{b}{}^{c} 
   \nonumber\\&\;\;
   + \tfrac{593}{105} \sigma _{a}{}^{d} \Theta   u^{b}  u^{c} \nabla _{c}\sigma _{bd} + \tfrac{28}{5} \sigma _{a}{}^{d} \sigma _{d}{}^{e}  u^{b}  u^{c} \nabla _{c}\sigma _{be} - \tfrac{1}{15} \Theta   u^{b} \nabla _{c}\nabla _{a}\sigma _{b}{}^{c} + \tfrac{2}{35} \nabla _{c}\nabla _{a}\nabla _{b}\sigma ^{bc}- \tfrac{124}{105} \Theta   u^{b} \nabla _{c}\nabla _{b}\sigma _{a}{}^{c} 
   \nonumber\\&\;\;
   +\tfrac{2}{35} \nabla _{c}\nabla _{b}\nabla _{a}\sigma ^{bc}+ \tfrac{2}{15}\nabla^{b}\nabla _{b}\nabla _{a}\Theta  + \tfrac{4}{5}  u^{b}  u^{c} \nabla _{c}\nabla _{b}\nabla _{a}\Theta  + \tfrac{2}{35} \nabla^{b}\nabla _{b}\nabla _{d}\sigma _{a}{}^{d} + \tfrac{16}{35}  u^{b}  u^{c} \nabla _{c}\nabla _{b}\nabla _{d}\sigma _{a}{}^{d} 
   \nonumber\\&\;\;
    + \tfrac{16}{35}  u^{b}  u^{c} \nabla _{c}\nabla _{d}\nabla _{b}\sigma _{a}{}^{d} - \tfrac{13}{105}  u^{b} \nabla _{a}\sigma _{bc} \nabla ^{c}\Theta  - \tfrac{148}{105}  u^{b} \nabla _{b}\sigma _{ac} \nabla ^{c}\Theta - \tfrac{2}{35}  u^{b} \nabla _{c}\sigma _{ab} \nabla ^{c}\Theta  + \tfrac{14}{5} \sigma _{b}{}^{d} \sigma ^{bc} \nabla _{d}\sigma _{ac} 
   \nonumber\\&\;\;
    - \tfrac{4}{21}  u^{b}  u^{c}  u^{d} \nabla _{a}\Theta  \nabla _{d}\sigma _{bc} - \tfrac{9}{35} u^{b} \nabla _{c}\sigma _{a}{}^{e} \nabla^{c}\sigma _{be} + \tfrac{14}{5} \sigma _{a}{}^{b} \sigma _{b}{}^{c} \nabla _{d}\sigma _{c}{}^{d} - \tfrac{9}{35}  u^{b} \nabla _{a}\sigma _{b}{}^{c} \nabla _{d}\sigma _{c}{}^{d} 
   \nonumber\\&\;\;
    - \tfrac{12}{7}  u^{b}  u^{c}  u^{d} \nabla _{a}\sigma _{b}{}^{e} \nabla _{d}\sigma _{ce} - \tfrac{174}{35}  u^{b}  u^{c}  u^{d} \nabla _{b}\sigma _{a}{}^{e} \nabla _{d}\sigma _{ce} - \tfrac{1}{7} \sigma ^{cd}  u^{b} \nabla _{d}\nabla _{a}\sigma _{bc} - \tfrac{23}{105} \Theta   u^{b}  u^{c}  u^{d} \nabla _{d}\nabla _{a}\sigma _{bc}
    \nonumber\\&\;\;
    - \tfrac{64}{35} \sigma _{a}{}^{c}  u^{b} \nabla _{d}\nabla _{b}\sigma _{c}{}^{d} - \tfrac{1}{15} \Theta   u^{b} \nabla^{c}\nabla _{c}\sigma _{ab} - \tfrac{26}{21} \Theta   u^{b}  u^{c}  u^{d} \nabla _{d}\nabla _{c}\sigma _{ab}  - \tfrac{78}{35} \sigma _{a}{}^{e}  u^{b}  u^{c}  u^{d} \nabla _{d}\nabla _{c}\sigma _{be} 
    \nonumber\\&\;\;
    + \tfrac{2}{35} \nabla _{d}\nabla^{b}\nabla _{b}\sigma _{a}{}^{d} + \tfrac{16}{35}  u^{b}  u^{c} \nabla _{d}\nabla _{c}\nabla _{b}\sigma _{a}{}^{d} - \tfrac{96}{35}  u^{b} \nabla _{b}\sigma _{cd} \nabla ^{d}\sigma _{a}{}^{c} - \tfrac{9}{35}  u^{b} \nabla _{a}\sigma _{cd} \nabla ^{d}\sigma _{b}{}^{c} + 4 \sigma _{a}{}^{b} \sigma ^{cd} \nabla _{d}\sigma _{bc} 
    \nonumber\\&\;\;
    - \tfrac{4}{3} \dot{\Theta } \nabla _{a}\Theta  - \tfrac{2}{35}  u^{b} \nabla _{a}\Theta  \nabla _{c}\sigma _{b}{}^{c} - \tfrac{9}{5} \Theta   u^{b} \nabla _{b}\nabla _{a}\Theta  + \tfrac{23}{7} \sigma _{ab} \Theta  \nabla ^{b}\Theta  - \tfrac{64}{35} \sigma ^{cd} \, u^{b} \nabla _{d}\nabla _{b}\sigma _{ac}
    \nonumber\\&\;\;
    - \tfrac{1}{7} \sigma _{a}{}^{e}  u^{b} \nabla^{c}\nabla _{c}\sigma _{be} + \tfrac{2}{35}\nabla^{b}\nabla _{d}\nabla _{b}\sigma _{a}{}^{d} - \tfrac{96}{35}  u^{b} \nabla _{b}\sigma _{a}{}^{c} \nabla _{d}\sigma _{c}{}^{d}\,,
\\   \nonumber\;\;& \\ 
   \Xbbd_{ ab } & \peq
   8 \sigma _{ a}{}^{c} \sigma _{b }{}^{d} \sigma _{c}{}^{e} \sigma _{de} + \tfrac{256}{21}\Theta \sigma _{ a}{}^{c} \sigma _{b }{}^{d} \sigma _{cd}   + \tfrac{10}{21}\Theta \sigma _{ ab } \sigma_{cd}\sigma^{cd}
    \nonumber\\&\;\; \nonumber
   + \tfrac{2}{21}\big(81 \Theta ^2-     76\dot{\Theta }\big) \sigma _{ a}{}^{c} \sigma _{b c}  + 
   \tfrac{2}{189}\Theta \big(251 \Theta^2 
   - 564  \dot{\Theta }\big)\sigma _{ ab }
   \nonumber\\&\;\; \nonumber
   + \tfrac{1}{63}\big(30\dot{\Theta } -43\Theta ^2\big)  u^{c} \nabla _{ a}\sigma _{b c}  - \tfrac{7}{3}\Theta \sigma _{ a}{}^{d}  u^{c} \nabla _{b }\sigma _{cd} + \tfrac{6}{7} \nabla _{ a}\sigma ^{cd} \nabla _{b }\sigma _{cd} + \tfrac{2}{21}  u^{c}  u^{d} \nabla _{ a}\Theta  \nabla _{b }\sigma _{cd} 
   \\&\;\; \nonumber
   + \tfrac{3}{7}  u^{c}  u^{d} \nabla _{ a}\sigma _{c}{}^{e} \nabla _{b }\sigma _{de} + \tfrac{2}{3} \nabla _{ a}\Theta  \nabla _{b }\Theta  + \tfrac{8}{7} \sigma ^{cd} \nabla _{ a}\nabla _{b}\sigma _{cd} + \tfrac{2}{21} \Theta   u^{c}  u^{d} \nabla _{ a}\nabla _{b}\sigma _{cd} + \tfrac{8}{9} \Theta  \nabla _{ a}\nabla _{b}\Theta  
   \\&\;\; \nonumber
   - \tfrac{2}{3}  u^{c} \nabla _{ a}\nabla _{c}\nabla _{b}\Theta  - \tfrac{2}{7}  u^{c} \nabla _{ a}\nabla _{c}\nabla _{d}\sigma _{b }{}^{d} + \tfrac{8}{7} \sigma ^{cd} \nabla _{ a}\nabla _{d}\sigma _{b c} + \tfrac{11}{21} \Theta   u^{c}  u^{d} \nabla _{ a}\nabla _{d}\sigma _{b c} + \tfrac{8}{7} \sigma _{ a}{}^{c} \nabla _{b }\nabla _{d}\sigma _{c}{}^{d}  
   \nonumber\\&\;\;
   - \tfrac{2}{7}  u^{c} \nabla _{ a}\nabla _{d}\nabla _{c}\sigma _{b }{}^{d} - \tfrac{34}{9} \Theta ^2  u^{c} \nabla _{c}\sigma _{ ab } + \tfrac{88}{21} \dot{\Theta }  u^{c} \nabla _{c}\sigma _{ ab } + \tfrac{15}{7}  u^{c}  u^{d} \nabla _{ a}\sigma _{de} \nabla _{c}\sigma _{b }{}^{e} + \tfrac{20}{21} \nabla _{ a}\Theta  \nabla _{c}\sigma _{b }{}^{c} 
   \nonumber\\&\;\;
   - \tfrac{118}{7} \sigma _{ a}{}^{d} \sigma _{d}{}^{e}  u^{c} \nabla _{c}\sigma _{b e} - \tfrac{86}{7} \sigma _{ a}{}^{d} \sigma _{b }{}^{e}  u^{c} \nabla _{c}\sigma _{de} + \tfrac{16}{7} \sigma _{ ab }  u^{c} \nabla _{c}\dot{\Theta } + \tfrac{16}{21} \Theta  \nabla _{c}\nabla _{ a}\sigma _{b }{}^{c} + \tfrac{8}{7} \sigma _{ a}{}^{c} \nabla _{c}\nabla _{b }\Theta  
   \nonumber\\&\;\;
   - \tfrac{2}{7}  u^{c} \nabla _{c}\nabla _{ a}\nabla _{d}\sigma _{b }{}^{d} - \tfrac{2}{7}  u^{c} \nabla _{c}\nabla _{d}\nabla _{ a}\sigma _{b }{}^{d} - \tfrac{1}{7}  u^{c} \nabla _{c}\nabla^d\nabla _{d}\sigma _{ ab } + \tfrac{20}{21} \nabla _{ a}\sigma _{b c} \nabla ^{c}\Theta  + \tfrac{10}{21} \nabla _{c}\sigma _{ ab } \nabla ^{c}\Theta 
   \nonumber\\&\;\;
   + \tfrac{6}{7} \nabla _{c}\sigma _{ a}{}^{d} \nabla^c\sigma _{b d} + \tfrac{60}{7}  u^{c}  u^{d} \nabla _{c}\sigma _{ a}{}^{e} \nabla _{d}\sigma _{b e} + \tfrac{12}{7} \nabla _{ a}\sigma _{b }{}^{c} \nabla _{d}\sigma _{c}{}^{d} + \tfrac{15}{7}  u^{c}  u^{d} \nabla _{ a}\sigma _{b }{}^{e} \nabla _{d}\sigma _{ce}
   \nonumber\\&\;\;
   + \tfrac{8}{7} \sigma _{ a}{}^{c} \nabla _{d}\nabla _{b }\sigma _{c}{}^{d} + \sigma _{ a}{}^{e}  u^{c}  u^{d} \nabla _{d}\nabla _{b }\sigma _{ce} - \tfrac{2}{7}  u^{c} \nabla _{d}\nabla _{ a}\nabla _{c}\sigma _{b }{}^{d} + \tfrac{8}{21} \Theta  \nabla^c\nabla _{c}\sigma _{ ab } + \tfrac{76}{21} \Theta   u^{c}  u^{d} \nabla _{d}\nabla _{c}\sigma _{ ab } 
   \nonumber\\&\;\;
   + \tfrac{80}{7} \sigma _{ a}{}^{e}  u^{c}  u^{d} \nabla _{d}\nabla _{c}\sigma _{b e} + \tfrac{2}{7} \sigma _{ ab } \nabla^c\nabla _{c}\Theta  - \tfrac{2}{7}  u^{c} \nabla _{d}\nabla _{c}\nabla _{ a}\sigma _{b }{}^{d} + \tfrac{12}{7} \nabla _{ a}\sigma _{cd} \nabla ^{d}\sigma _{b }{}^{c} - \tfrac{1}{7} u^{c} \nabla^d\nabla _{c}\nabla _{d}\sigma _{ ab } 
   \nonumber\\&\;\;
   - \tfrac{10}{7}  u^{c}  u^{d}  u^{e} \nabla _{e}\nabla _{d}\nabla _{c}\sigma _{ ab }   -2 \sigma _{ a}{}^{d} \sigma _{d}{}^{e}  u^{c} \nabla _{b }\sigma _{ce} + \tfrac{16}{21} \Theta  \nabla _{ a}\nabla _{c}\sigma _{b }{}^{c} + \sigma _{ a}{}^{e}  u^{c}  u^{d} \nabla _{b }\nabla _{d}\sigma _{ce}
   \nonumber\\&\;\;
   - \tfrac{428}{21} \sigma _{ a}{}^{d} \Theta   u^{c} \nabla _{c}\sigma _{b d} - \tfrac{1}{3}  u^{c} \nabla _{c}\nabla _{ a}\nabla _{b }\Theta  + \tfrac{10}{21}  u^{c}  u^{d} \nabla _{ a}\Theta  \nabla _{d}\sigma _{b c} + \tfrac{11}{21} \Theta   u^{c}  u^{d} \nabla _{d}\nabla _{ a}\sigma _{b c} 
   \nonumber\\&\;\;
   - \tfrac{1}{7} u^{c} \nabla^d\nabla _{d}\nabla _{c}\sigma _{ ab }  + \tfrac{8}{7} \sigma ^{cd} \nabla _{d}\nabla _{ a}\sigma _{b c} + \tfrac{8}{7}\sigma _{ a}{}^{e} \nabla^c\nabla _{c}\sigma _{b e}\,,
\\ \nonumber\;\;& \\
   \Xbbd_{ abc} &  \peq 
   \tfrac{166}{27} \sigma _{ b}{}^{d} \sigma _{cd} \nabla _{a }\Theta  + \tfrac{464}{81}\Theta  \sigma _{ ab}  \nabla _{c }\Theta  
   + \tfrac{452}{27} \Theta \sigma _{ a}{}^{d}  \nabla _{c}\sigma _{b d} 
   + \tfrac{2}{81}\big(134 \Theta ^2  - 141 \dot{\Theta }\big) \nabla _{ c}\sigma _{ab } 
   \nonumber \\&\;\; 
   + 10 \sigma _{ a}{}^{d} \sigma _{b}{}^{e} \nabla _{c }\sigma _{de} -  u^{d} \nabla _{ b}\sigma _{a}{}^{e} \nabla _{c }\sigma _{de} - \tfrac{7}{27} \Theta   u^{d} \nabla _{ c}\nabla _{b}\sigma _{a d} - \tfrac{5}{9} \sigma _{ a}{}^{e}  u^{d} \nabla _{c}\nabla _{b }\sigma _{de} + \tfrac{1}{3} \nabla _{ a}\nabla _{b}\nabla _{c }\Theta  
   \nonumber\\&\;\; \nonumber
   - \tfrac{26}{9} \Theta   u^{d} \nabla _{ c}\nabla _{d}\sigma _{ab } - \tfrac{80}{9} \sigma _{ a}{}^{e}  u^{d} \nabla _{c}\nabla _{d}\sigma _{b e} + \tfrac{2}{9} \nabla _{ c}\nabla _{d}\nabla _{b}\sigma _{a }{}^{d} + \tfrac{1}{9}\nabla _{ c}\nabla^d\nabla _{d}\sigma _{ab } + \tfrac{10}{9}  u^{d}  u^{e} \nabla _{ c}\nabla _{e}\nabla _{d}\sigma _{ab } 
   \\&\;\; \nonumber
   - \tfrac{94}{27}  u^{d} \nabla _{ a}\Theta  \nabla _{d}\sigma _{bc } + \tfrac{4}{9} \Theta \sigma _{ ab}  \nabla _{d}\sigma _{c }{}^{d} - \tfrac{40}{3}  u^{d} \nabla _{ b}\sigma _{a}{}^{e} \nabla _{d}\sigma _{c e} - \tfrac{26}{9} \Theta   u^{d} \nabla _{d}\nabla _{ c}\sigma _{ab } - \tfrac{80}{9} \sigma _{ a}{}^{e}  u^{d} \nabla _{d}\nabla _{c}\sigma _{b e} 
   \\&\;\; 
   + \tfrac{2}{9} \nabla _{d}\nabla _{ c}\nabla _{b}\sigma _{a }{}^{d} + \tfrac{8}{9} \sigma _{ ab} \sigma _{c d} \nabla ^{d}\Theta  + \tfrac{1}{9}\nabla^d\nabla _{ c}\nabla _{d}\sigma _{ab } + \tfrac{10}{9}  u^{d}  u^{e} \nabla _{e}\nabla _{ c}\nabla _{d}\sigma _{ab } + \tfrac{1}{9}\nabla^d\nabla _{d}\nabla _{ c}\sigma _{ab } 
   \nonumber\\&\;\;
   + 14 \sigma _{ a}{}^{d} \sigma _{d}{}^{e} \nabla _{c}\sigma _{b e} + \tfrac{2}{9} \nabla _{ c}\nabla _{b}\nabla _{d}\sigma _{a }{}^{d}
   + \tfrac{4}{9} \Theta \sigma _{ a}{}^{d}  \nabla _{d}\sigma _{bc } -4 \sigma _{ ab}  u^{d} \nabla _{d}\nabla _{c }\Theta  + \tfrac{10}{9}  u^{d}  u^{e} \nabla _{e}\nabla _{d}\nabla _{ c}\sigma _{ab }
    \nonumber\\&\;\;
   - \tfrac{2}{9}  u^{d} \nabla _{ a}\Theta  \nabla _{c}\sigma _{b d} \,,
\\ \nonumber\;\;& \\
   \Xbbd_{ abcd } & \peq 
   \tfrac{10}{3} \Theta \sigma _{ ab} \sigma _{c}{}^{e} \sigma _{d e}  + \tfrac{2}{9}\big(13 \Theta ^2  -9\dot{\Theta }\big) \sigma _{ ab} \sigma _{cd }  + \tfrac{10}{3} \nabla _{ a}\Theta  \nabla _{d}\sigma _{bc } + 6 \nabla _{ b}\sigma _{a}{}^{e} \nabla _{d}\sigma _{c e} 
   \nonumber \\&\;\; 
   + 2 \sigma _{ ab} \nabla _{d}\nabla _{c }\Theta  -  u^{e} \nabla _{ d}\nabla _{c}\nabla _{e}\sigma _{ab } - \, u^{e} \nabla _{ d}\nabla _{e}\nabla _{c}\sigma _{ab }  -  u^{e} \nabla _{e}\nabla _{ d}\nabla _{c}\sigma _{ab }
   -2 \Theta \sigma _{ ab} \dot\sigma _{cd} 
   \nonumber\\&\;\; 
   + 8 \sigma _{ a}{}^{e} \nabla _{d}\nabla _{c}\sigma _{b e} + \tfrac{8}{3} \Theta  \nabla _{ d}\nabla _{c}\sigma _{ab } \,,
  \\ \nonumber\;\;&
    \\
%###############################################################################
   \Xbbd_{ abcde} & \peq
   2 \sigma _{ bc} \sigma _{de} \nabla _{a}\Theta  + 2\Theta \sigma _{ ab}   \nabla _{e}\sigma _{cd} + \nabla _{ e}\nabla _{d}\nabla _{c}\sigma _{ab} \,.
\end{align}
\endgroup
\vspace{0.1cm}
%\newpage

\subsection*{Multipoles of $\mathbb{Y}^{(1,2)}$} 

\vspace{0.1cm}

\begin{align}
  \mathbb{Y}^{(1)} = \langle\mathbb{Y}^{(1)}\rangle + \mathbb{Y}^{(1)}_a \;n^a + \mathbb{Y}^{(1)}_{\langle ab\rangle}\;n^{a}n^{b} \,, 
\end{align}
where
\begin{align}
\langle\mathbb{Y}^{(1)}\rangle &=\tfrac{1}{3}R^a_a+\tfrac{4}{3}u^a u^b R_{ab}\,,
\\
\mathbb{Y}^{(1)}_a & \peq -2u^b R_{ab}\,,
\\
\mathbb{Y}^{(1)}_{ ab} & \peq R_{ ab }\,.
\end{align}

\begin{equation}
    \mathbb{Y}^{(2)} = \langle\mathbb{Y}^{(2)}\rangle + \mathbb{Y}^{(2)}_a \;n^a + \mathbb{Y}^{(2)}_{\langle ab\rangle}\;n^{a}n^{b} + \mathbb{Y}^{(2)}_{\langle abc\rangle }\;n^{a}n^bn^{c}\;,
\end{equation}
where 
\begin{align}
\langle\mathbb{Y}^{(2)}\rangle &= - \tfrac{1}{3}  \dot R - \tfrac{2}{3} 
u^{a} \nabla _{b} R_{a}{}^{b}  -2  u^{a}  u^{b}  \dot R_{ab}\,,
\\
\mathbb{Y}^{(2)}_a & \peq \tfrac{6}{5}  u^{b}  u^{c} \nabla _ {a} R_{bc} + \tfrac{1}{5} \
\nabla _ {a} R + \tfrac{2}{5} \nabla _ {b}
R_{a}{}^{b} + \tfrac{12}{5}  u^{b}  \dot R_{ab}\,,
\\
\mathbb{Y}^{(2)}_{ab} & \peq -2  u^{c} \nabla _ {b} R_{a c} -  \dot R_{ab }\,,
\\
\mathbb{Y}^{(2)}_{abc} & \peq \nabla _ {a} R_{bc}\,.
\end{align}

\newpage

\section{Multipoles of the expansion rate fluctuation field}
\label{app_eta_Xl}

We present the relation between the multipoles of the expansion rate fluctuation field $\eta_\ell$  and the dominant  multipoles of the covariant cosmographic parameters in the axial symmetric configuration

\allowdisplaybreaks
\begingroup
\begin{align}\label{dipe}
\eta_1(z) &\approx \frac{\mathbb{Q}_1}{2 \ln10}\, z-\frac{(9 \Qbb_0+5) \Qbb_1 z^2}{12 \ln10}-\frac{1}{27720 \ln10}\bigg\{-33 \Qbb_1 \bigg[280 \Jbb_0+112 \Jbb_2
\\ \notag
&~~
-5 \left(14 \Qbb_0 (15 \Qbb_0+16)+46 \Qbb_3^2+28 \Rbb_0+63\right)\bigg]+8 \Qbb_3 \left(-297
   \Jbb_2-220 \Jbb_4+135 \Qbb_3^2\right)
\\ \notag
&~~
+6930 \Qbb_1^3+5940 \Qbb_1^2 \Qbb_3-1155 \Sbb_1\bigg\}\,z^3 \,,
\\ \notag & \\
%\end{align}
%\begin{align}
\label{quade}  
 \eta_2(z)&\approx \frac{\mathbb{H}_2}{\mathbb{H}_0\ln10}+
\frac{14 \Jbb_2-3 (\Qbb_1+\Qbb_3) (7 \Qbb_1+2 \Qbb_3)}{84 \ln10}z^2+\frac{1}{504
   \ln10}\bigg[-21 \Jbb_2 (8 \Qbb_0+7)
\\ \notag
&~~
+60 \Qbb_0 (\Qbb_1+\Qbb_3) (7 \Qbb_1+2 \Qbb_3)+224 \Qbb_1^2+4 \Qbb_3 (81 \Qbb_1+20 \Qbb_3)\bigg]\,z^3 \,,
\\ \notag & \\
%\end{align}
%\begin{align}
\label{octe}
\eta_3(z) &\approx \frac{\mathbb{Q}_3}{2 \ln10} z-\frac{(9 \Qbb_0+7) \Qbb_3 z^2}{12 \ln10}
\\ \notag
&~~
+ \frac{1}{154440 \ln10}\bigg[-6435 (\Qbb_3 (8 \Jbb_0-6 \Qbb_0 (5 \Qbb_0+6)-13)+\Sbb_3)
\\ \notag
&~~
-3432 \Jbb_2 (9 \Qbb_1+4 \Qbb_3)-1040 \Jbb_4 (22 \Qbb_1+9 \Qbb_3)
\\ \notag
&~~
+30
   \left(858 \Qbb_1^3+3289 \Qbb_1^2 \Qbb_3+1404 \Qbb_1 \Qbb_3^2+723 \Qbb_3^3+858 \Qbb_3 \Rbb_0\right)\bigg]\,z^3 \,,
\\ \notag & \\
%\end{align}
%\begin{align}
\label{hexe}  
 \eta_4(z)&\approx
\frac{154 \Jbb_4-9 \Qbb_3 (44 \Qbb_1+9 \Qbb_3)}{924 \ln10}z^2
\\ \notag
&~~
+\frac{\Big[12 \Qbb_3 (44 (5 \Qbb_0+3) \Qbb_1+15 (3 \Qbb_0+2) \Qbb_3)-77 \Jbb_4 (8 \Qbb_0+7)\Big]}{1848 \ln10}\,z^3 \,,
\\ \notag & \\
%\end{align}
%\begin{align}\label{p32}  
 \eta_5(z)&\approx
\frac{\left\{60 \Qbb_3 \Big[5 \left(26 \Qbb_1^2+27 \Qbb_1 \Qbb_3+6 \Qbb_3^2\right)-52 \Jbb_2\Big]-40 \Jbb_4 (91 \Qbb_1+36 \Qbb_3)-819
   \Sbb_5\right\}}{19656 \ln10}\,z^3 \,,
\end{align}
and for the monopole
\begin{align}\label{monom}
\mathcal{M}(z)&\approx\log \mathbb{H}_0-\frac{1-\mathbb{Q}_0}{2\ln10}\, z +\frac{\left(28 \Jbb_0-7 \Qbb_0 (9 \Qbb_0+10)-21 \Qbb_1^2-9 \Qbb_3^2-28 \Rbb_0+49\right)}{168 \ln10}\, z^2
\\ \notag
&~~
+\frac{1}{504 \ln10}\bigg[-21 \Jbb_0 (8 \Qbb_0+7)+3 \Qbb_0 \left(14 \Qbb_0 (5 \Qbb_0+8)+70 \Qbb_1^2+30 \Qbb_3^2+28 \Rbb_0+63\right)
\\ \notag
&~~
+112 \Qbb_1^2+60 \Qbb_3^2+84 \Rbb_0-21 \Sbb_0-105\bigg]\,z^3 \,.
\end{align}
\endgroup

\newpage
\bibliographystyle{JHEP}
%\bibliography{bib}

\providecommand{\href}[2]{#2}\begingroup\raggedright
\endgroup

\end{document}